\documentclass[prb,showpacs,preprintnumbers,amsmath,amssymb, superscriptaddress]{revtex4}
\usepackage{graphicx}
\usepackage{ifpdf}
\usepackage{subfigure}
\usepackage{dcolumn}
\usepackage{bm}

\begin{document}
\preprint{APS/123-QED}
\title{Takagi-Taupin Description of X-ray Dynamical Diffraction from Diffractive Optics with Large Numerical Aperture}
\author{Hanfei Yan}
\email[Electronic mail: ]{hanfei@aps.anl.gov}
\author{Jorg Maser}
\affiliation{Center for Nanoscale Materials, Argonne National Laboratory, Argonne, IL 60439}
\affiliation{X-ray Science Division, Argonne National Laboratory, Argonne, IL 60439}
\author{Albert Macrander}
\author{Qun Shen}
\author{Stefan Vogt}
\affiliation{X-ray Science Division, Argonne National Laboratory, Argonne, IL 60439}
\author{Brian Stephenson}
\affiliation{Materials Science Division, Argonne National Laboratory, Argonne, IL 60439}
\affiliation{Center for Nanoscale Materials, Argonne National Laboratory, Argonne, IL 60439}
\author{Hyon Chol Kang}
\affiliation{Materials Science Division, Argonne National Laboratory, Argonne, IL 60439}
\affiliation{Center for Nanoscale Materials, Argonne National Laboratory, Argonne, IL 60439}
\affiliation{Advanced Photonics Research Institute,Gwangju Institute of Science and Technology,Gwangju, 500-712, Republic of Korea}

\date{\today}
\begin{abstract}
We present a formalism of x-ray dynamical diffraction from volume diffractive optics with large numerical aperture and high aspect ratio, in an analogy to the Takagi-Taupin equations for strained single crystals. We derive a set of basic equations for dynamical diffraction from volume diffractive optics, which enable us to study the focusing property of these optics with various grating profiles. We study volume diffractive optics that satisfy the Bragg condition to various degrees, namely flat, tilted and wedged geometries, and derive the curved geometries required for ultimate focusing. We show that the curved geometries satisfy the Bragg condition everywhere and phase requirement for point focusing, and effectively focus hard x-rays to a scale close to the wavelength.
\end{abstract}

\maketitle

\section{Introduction}
X-ray microscopy has found numerous applications in materials sciences, medicine, biology, environmental sciences and many other fields by utilizing analytical techniques such as diffraction, x-ray fluorescence and spectroscopy, as well as imaging  techniques using absorption and phase contrast. Progress in the fabrication of x-ray optics has pushed the spatial resolution of hard x-ray microscopes and -microprobes to well below 100 nm using Kirk Patrick-Baez (K-B) mirrors\cite{Liu2005a,Mimura2005,Mimura2007,Yumoto2005}, refractive lenses\cite{Schroer2005} and diffractive optics\cite{Kang2006,Suzuki2005}. This makes hard x-ray microscopy and spectromicroscopy powerful tools to probe the structure and properties of materials at the nanoscale, with the advantages of good penetration, which provides for the study of thick specimens and buried structures. As well, there are only moderate or no vacuum requirements, enabling in-situ studies, and one can avoid interaction with electric and magnetic fields, enabling good quantification. In order to push the spatial resolution of x-ray microscopy to the theoretical limit, it is important to know whether there exists a limit in effectively focusing x-rays using optics made of realistic materials. Such limits have been studied theoretically for various optics, and yielded numbers of the resolution limit of approximately 10 nm for waveguides\cite{Bergemann2003} and 2 nm or less for a refractive lens in an idealized adiabatically focusing geometry\cite{Schroer2005a}. For diffractive optics, for example a zone plate with 'flat zones', a recent study found that when illuminated by a plane wave a zone plate could have a similar focusing limit as that for a waveguide, while for 1:1 imaging (spherical wave illumination), where a Bragg condition is satisfied, no limit to the spatial resolution was found\cite{Pfeiffer2006}.
Diffractive optics are intrinsically well suited to achieve a high spatial resolution because a large numerical aperture (NA) can be achieved by diffraction. A novel approach to making diffractive optics with high NA is the Multilayer-Laue-Lens (MLL)\cite{Liu2005}. MLL's are fabricated by multilayer deposition onto a flat substrate and used in Laue geometry. An MLL can be considered as a special type of zone plate. The fabrication approach provides high aspect ratios, and is therefore well suited for hard x-ray focusing\cite{Maser2004}. Initial 1-D MLL structures have achieved a line focus of 30 nm in the hard x-ray range (19.5 keV) with a diffraction efficiency of more than 40\%\cite{Kang2006}. 

MLL's exhibit strong dynamic diffraction properties, and have to be described using a dynamical diffraction theory. Using such an approach, namely Coupled-Wave-Theory (CWT)\cite{Solymar1981}, we have shown that MLL's can reach a spatial resolution of one nanometer\cite{Kang2006}. However, these simulation results were limited by assumptions made in the derivation of the local one-dimensional (1-D) CWT equations for volume diffractive optics\cite{Maser1992}. To understand the ultimate limit of focusing by diffractive x-ray optics made of realistic materials, we have developed a new theoretical approach to volume diffraction, based on the Takagi-Taupin description of dynamical x-ray diffraction developed for strained single crystals. This approach represents a full-wave dynamical theory for volume diffractive optics with high numerical aperture. It takes into account flat, tilted, wedged and curved zones as well as arbitrary zone profiles, allowing us to simulate any of these geometries for imaging and focusing. Since it is based on a dynamical diffraction theory, it provides physical insight into the diffraction properties of volume diffractive optics. Using this approach, we have studied the focusing performance of various MLL's, and have investigated limitations that constrain the achievable NA. We find that a wedged MLL is able to focus hard x-rays to less tahn a nanometer, while curved zones are needed to achieve a beam size approaching the x-ray wavelength.

\section{THEORETICAL APPROACH}
In the conventional geometrical-optical theory, a zone plate is considered to be ``thin'' so that the volume diffraction effect can be neglected\cite{Kirtz1974}. This approximation is only valid when the zone plate thickness, $w$, is less than $(2\Delta r_{min})^2/\lambda$, where $\Delta r_{min}$ is the outmost zone width and $\lambda$ is the wave length of the incident x-ray. Such a condition is usually satisfied for diffractive optics at optimum thickness with a zone width no smaller than 10 nm\cite{Maser2004}. A more rigorous full-wave theory that takes into account the dynamical diffraction effect is needed to correctly describe x-ray diffraction from a zone plate with outmost zone width less than 10 nm because the optimized thicknesses are too large to treat them as ``thin''. This was initially done by Maser and Schmahl\cite{Maser1992}, who applied locally a 1-D approach of CWT to study the variation of the local diffraction efficiency with thickness and with the slant of the zones. Their approach assumes that the zone plate can be decomposed locally into periodic gratings with fixed gratings period. Their approach properly accounts for dynamical diffraction effects, and allows the calculation of the point spread function and modulation transfer function of zone plates with high aspect ratios and small outermost zone width. However, their approach is limited to cases of  $w<<f$ and relatively small NA, where $f$ is the focal length of the zone plate. For an optimized zone plate thickness (these are less than $1\ \mu$m for soft x-rays and greater than $1\ \mu$m for hard x-rays), and for a zone plate radius matched to the coherence length of a $3^{rd}$ generation synchrotron source, their approach is limited to MLL's with outmost zone width no smaller than 1 nm. 

Other calculations have been published since that of Maser and Schmahl. The 1-D CWT approach was further extended to study the effects of an arbitrary line/space ratio and interface roughness by Schneider\cite{Schneider1998,Schneider1997}. By decomposing zone plates into local gratings, and with similar limitations as 1-D CWT, Levashov and Vinogradov\cite{Levashov1994} studied the variation of the total diffraction efficiency with thickness. A different numerical approach based on the parabolic wave equation within paraxial approximation has been employed by Kurokhtin and Popov\cite{Kurokhtin2002}. Their numerical method requires large computing power when thousands of individual zones are considered. More recently methods of solving the parabolic wave equation using eigenfunctions have been reported\cite{Pfeiffer2006,Schroer2006}. Due to the use of the paraxial approximation, numerical approaches based on the parabolic wave equation are limited to intermediate values of numerical aperture (NA). To overcome the limitations of these theories, and to provide an approach that is valid to wavelength scale, we have developed and present below a new approach to modeling dynamical diffraction from MLL's based on first principles. 

From Maxwell's equations, we obtain the scalar-wave equation describing the electric field variation of a monochromatic x-ray wave in a medium with susceptibility $\chi(\vec{r})$ as,
\begin{equation}
	\nabla^{2}\vec{E}(\vec{r})+k^{2}[1+\chi(\vec{r})]\vec{E}(\vec{r})=0
	\label{eq1}
\end{equation}
where $k=2\pi/\lambda$ and $\vec{E}$ is the electric field vector. In principle, for a given function $\chi(\vec{r})$, the electric field $\vec{E}$ can be solved directly from Eq. (\ref{eq1}). However, in many cases the variation of $\chi(\vec{r})$ is too complicated to allow an analytical solution to Eq. (1). Even using a numerical approach, the computing time can be prohibitive. An extreme case is x-ray diffraction from single crystals, where tens of thousands of atomic planes are involved in the scattering process and solving Eq. (\ref{eq1}) plane by plane becomes impractical. The same difficulty arises for x-ray diffraction from volume diffractive optics, which contains thousands of diffractive structures, and requires the consideration of the multiwave scattering process. In crystallography, this difficulty has been overcome elegantly by the classical dynamical diffraction theory, in which the periodic susceptibility function $\chi$ is expanded into a Fourier series. Each Fourier expansion coefficient, $\chi_{h}$, serves as a resonator, interacting with the incident plane wave and exciting a resonant wave. The excited resonant wave will again interact with resonators and excite high order resonant waves, leading to a dynamical multiwave scattering process and a self-consistent system of equations. Instead of seeking a solution to the second-order partial differential equation [Eq. (\ref{eq1})] for each atomic plane, the diffraction problem in a single crystal is simplified to finding a set of plane waves that satisfy the self-consistent system, thus allow a much simplified mathematical and numerical solution. To deal with strained single crystals, this approach was generalized by Takagi and Taupin\cite{Takagi1962,Taupin1964}. An additional phase term is added to each Fourier component of the susceptibility function to account for the phase change caused by the deformation of the lattice planes due to misorientation and strain fields. Detailed treatments of x-ray dynamical diffraction theory can be found in many monographs\cite{Authier2002,Pinsker1978}. We show that a similar methodology can be applied to solve the diffraction problem for diffractive optics.

We begin by considering a Fresnel Zone Plate (FZP). A FZP is a circular diffraction grating that is capable of focusing an incident wave to a series of focal points. It consists of alternating zones made of two different materials, and, in the limit of a 'thin' zone plate, produces a phase shift $\pi$ between waves passing through neighboring zones, thereby yielding a focusing effect. The position of the nth zone is determined by the zone plate law,
\begin{equation}
	r_{n}^{2}=n\lambda f+n^{2}\lambda^{2}/4,
	\label{eq2}
\end{equation}
where the width of the nth zone is given by,
\begin{equation}
	\Delta r_{n}=\frac{\lambda f}{2r_{n}}\sqrt{1+\frac{r_{n}^{2}}{f^{2}}}.
	\label{eq3}
\end{equation}
It is seen from Eq. (\ref{eq3}) that the zone width $\Delta r_{n}$ decreases radially. Locally, the zone width does not vary rapidly. When $r_{n}$ is large and $\Delta r_{n}$ is small, a FZP might be viewed as a ``strained crystal'' with $d$-spacing equal to $d=2\Delta r_{n}$. Therefore, we reason that a formalism akin to the Takagi-Taupin equations in crystallography can serve to allow a study of dynamical diffraction effects in a volume zone plate or MLL, and, more generally, in volume diffractive optics. A difficulty arises when applying the Takagi-Taupin equations since the $d$-spacing of volume diffractive optics changes significantly from its center to its outmost region. Its susceptibility function can therefore not be expanded into a Fourier series in a conventional way. We address this by finding a mapping relationship between zones in a FZP and zones in a strictly periodic structure with a known Fourier series expansion. From this mapping relationship, we expand the susceptibility function for a FZP into a pseudo Fourier series.

\begin{figure}[tbp]
	\centering
		\includegraphics[width=3.2in]{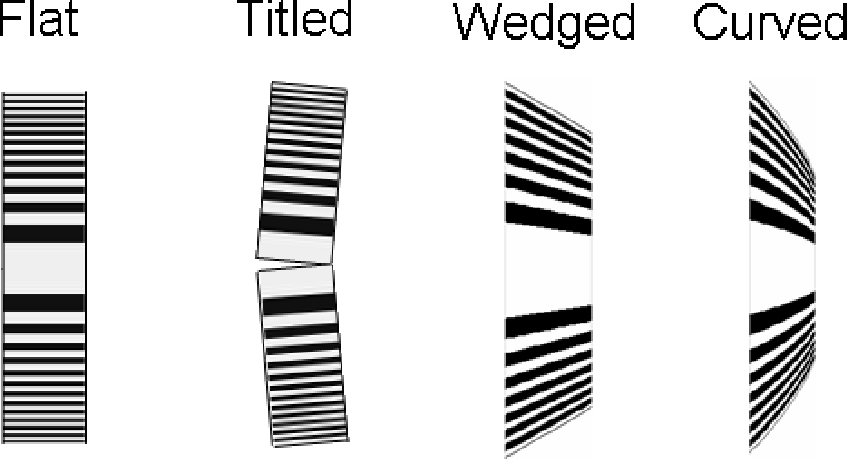}
	\caption{Types of multilayer-Laue-lens (MLL) layer arrangements:  flat, tilted, wedged, and curved.}
	\label{fig:FIG1}
\end{figure}

\begin{figure}[tbp]
	\centering
		\includegraphics[width=3.2in]{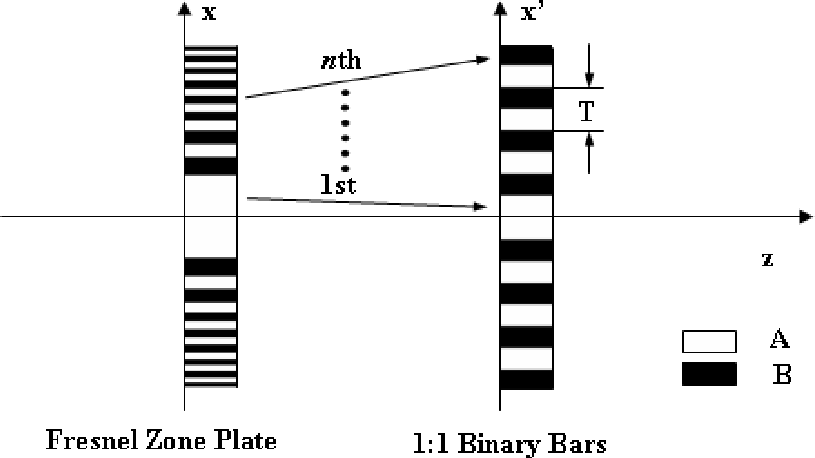}
	\caption{A mapping of layers between a Fresnel zone plate and a periodic structure composed by binary bars with thickness ratio 1:1.}
	\label{fig:FIG2}
\end{figure}

We will study in this paper flat, tilted, wedged and curved geometries (see Fig. \ref{fig:FIG1}). For the sake of simplicity, we begin by considering a 1-D MLL with flat zones. These zones can be aligned parallel to the optical axis in `flat geometry', satisfying the Bragg condition on the optical axis, or in tilted geometry, satisfying the Bragg condition in some outer area of the MLL. The zones are made of materials A and B, whose layer thickness follows the zone plate law Eq. (\ref{eq2}). This geometry has been manufactured as MLL's with outermost zone widths of 15 nm, 10 nm and 5 nm, and demonstrated to focus hard x-rays efficiently into a line with a width smaller than 20 nm\cite{KangUnpublished}. The aspect ratios of the manufactured MLL structures range from 1000 - 3000, more than an order of magnitude larger than what has been achieved using lithographic techniques. To find the pseudo Fourier series of an MLL, we consider another periodic binary structure with mirror symmetry about the axis $z$ as shown in Fig. \ref{fig:FIG2}. We consider that it is made of the materials A and B and has a A/B layer ratio of 1:1. This periodic grating has a radial coordinate, $x'$, that is different from the radial coordinate, $x$, for the MLL. A connection between $x$ and $x'$ will be established later. The Fourier series of a periodic structure can be written straightforwardly as,
\begin{eqnarray}
	\chi_{P}(x')&=&\bar{\chi}+\sum_{h=1}^{\infty}\frac{\Delta \chi}{h\pi}[1-(-1)^{h}]\sin(2h\pi\frac{\left|x'\right|}{T})\nonumber\\
						 &=&\bar{\chi}+\sum^{\infty}_{h=-\infty,h\neq 0}\frac{\Delta\chi}{2ih\pi}[1-(-1)^{\left|h\right|}]\exp(2h\pi\frac{\left|x'\right|}{T}),\ x'\neq 0
	\label{eq4}
\end{eqnarray}
where $\bar{\chi}=(\chi_{A}+\chi_{B})/2$ , $\Delta\chi=\chi_{A}-\chi_{B}$ and $T$ is the period. Here $T$ is taken as arbitrary. If a relationship between $x$ and $x'$ can be found so that the $n$th layer in the perfect periodic structure is mapped to the $n$th layer in a 1-D MLL, a series representation of the susceptibility function for the MLL can be obtained as, $\chi_{MLL}(x)=\chi_{P}[x'(x)]$. From the zone plate law [Eq. (\ref{eq2})], the $n$th layer index, $n$, can be written as,
	\[n=\frac{2}{\lambda}(\sqrt{x^{2}_{n}+f^{2}}-f).
\]
Similarly, in the periodic structure the $n$th layer index, $n'$, can be written as, 
	\[n'=\frac{2\left|x'_{n}\right|}{T}.
\]
By equating $n$ and $n'$ we establish a relationship between $x$ and $x'$ given by,
	\[\left|x'\right|=\frac{T}{\lambda}(\sqrt{x^{2}+f^{2}}-f).
\]
Substituting this into Eq. (\ref{eq4}) we obtain the following series expansion, 
\begin{equation}
	\chi_{MLL}(x)=\chi_{0}+\sum^{\infty}_{h=-\infty,h\neq 0}\chi_{h}\exp[i\phi_{h}(x)],\ x\neq 0,
	\label{eq5}
\end{equation}
where $\chi_{0}=\bar{\chi},\ \chi_{h}=(\Delta\chi/2ih\pi)[1-(-1)^{\left|h\right|}],\text{ and } \phi_{h}=hk(\sqrt{x^{2}+f^{2}}-f)$. Since the Fourier series of the periodic structure is mathematically complete and $x'$ is one-to-one mapped to $x$, the series expansion in Eq. (\ref{eq5}) can be considered as a pseudo Fourier series. Using Eq. (\ref{eq5}), we can construct the MLL structure, aside from the singular point at origin. For simplicity in the following discussion the subscript notation of MLL of $\chi_{MLL}$ will be dropped. 

\begin{figure}[tbp]
	\centering
		\includegraphics[width=3.2in]{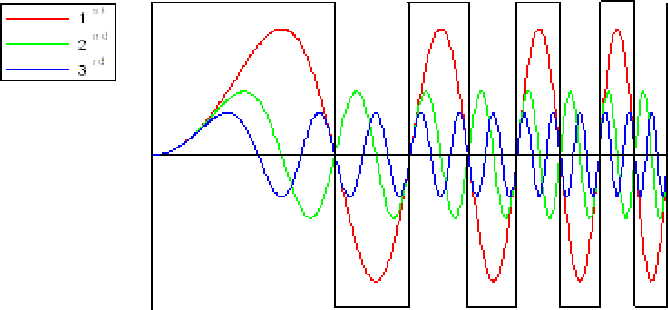}
	\caption{The zone plate structure can be represented by a set of sine waves with varying period that obey zone plate's law.}
	\label{fig:FIG3}
\end{figure}

We note that the gradient of the phase function, $\nabla\phi_{h}$, is equal to the local reciprocal lattice vector, $\vec{\rho}_{h}=(2\pi h/d)\vec{e}_{x}$, where $\vec{e}_{x}$ is a unit vector along the positive $x$ axis. Unlike the conventional Fourier series representation of a strictly periodic structure, here the zone plate structure is represented by a superposition of sine waves with varying periods obeying the zone plate law (see Fig. \ref{fig:FIG3}). If the radius of an MLL is sufficiently small so that $x_{max}$ is less than $(4f^{3}\lambda/\left|h\right|)^{1/4}$, the following approximation can be made,
\begin{equation}
	x^{2}_{n}\approx n\lambda f,\ \text{and $\phi_{h}\approx \frac{h\pi}{\lambda f}x^{2}$.}
	\label{eq6}
\end{equation}
The basic zone plate law is thereby reduced to the first term in Eq. (\ref{eq2}) only.

For an incident wave with an arbitrary wave front, we assume that it can be written as,
\begin{equation}
	\vec{E}_{incident}=\vec{E}^{(a)}_{0}(\vec{r})\exp(ik\vec{s}_{0}\cdot\vec{r}),
	\label{eq7}
\end{equation}
where $\vec{s}_{0}$ is a unit vector along the incident propagation direction, and $\vec{E}^{(a)}_{0}$ is a slowly varying complex function that represents the modulated amplitude of the incident wave. When an MLL is illuminated by such an incident wave, we write a trial solution to the wave equation as,
\begin{equation}
	\vec{E}=\sum_{h}{\vec{E}_{h}(\vec{r})\exp[i(k\vec{s}_{0}\cdot\vec{r}+\phi_{h})]}=\sum_{h}{\vec{E}_{h}P_{h}},
	\label{eq8}
\end{equation}
where $P_{h}=\exp[i(k\vec{s}_{0}\cdot\vec{r}+\phi_{h})]$.
This series form of the trial solution is chosen in accordance with the zone plate structure and Eq. (\ref{eq5}). Because a zone plate structure can be represented by a pseudo Fourier series, the phase term $P_{h}$ in the trial solution in Eq. (\ref{eq8}) indeed reflects the primary component of the spatial variation of the wavefield inside the global structure. Recall that $\vec{\rho}_{h}=\nabla\phi_{h}$, so in a local region the trial solution in Eq. (\ref{eq8}) is a Bloch wave solution. In addition, a further examination of $P_{h}$ shows that it actually corresponds to the phase variation of a spherical wave that converges to the $h$th order focus of the MLL, therefore this trial solution represents a superposition of a set of spherical waves with modulated amplitudes that converge to MLL's foci.

Substituting Eqs. (\ref{eq5}) and (\ref{eq8}) into Eq. (\ref{eq1}) we obtain,
\begin{eqnarray*}
\sum_{h}\sum_{i=x,y,z}[2\nabla E_{hi}\cdot\nabla P_{h}+P_{h}\nabla^{2}E_{hi}+E_{hi}(k^{2}P_{h}+\nabla^{2}P_{h})]\vec{e}_{i}
=-k^{2}\sum_{h}P_{h}(\sum_{l}\chi_{h-l}\vec{E}_{l}),
\end{eqnarray*}
where $\vec{e}_{x,y,z}$ are unit vectors along $x$, $y$ and $z$ axis. By equating them for each $h$, we obtain an infinite set of equations,
\begin{eqnarray}
&&\sum_{i=x,y,z}{[2\nabla E_{hi}\cdot\nabla P_{h}+P_{h}\nabla^{2}E_{hi}+E_{hi}(k^{2}P_{h}+\nabla^{2}P_{h})]\vec{e}_{i}}
=-k^{2}P_{h}\sum_{l}\chi_{h-l}\vec{E}_{l},\\
&&h,l=0,\pm 1,\pm 2, \pm 3,\ldots\nonumber
	\label{eq9}
\end{eqnarray}
If there is a solution to Eq. (\ref{eq9}), it will be also a solution to Eq. (\ref{eq1}). Eq. (\ref{eq9}) can be simplified using the following assumptions. First, the second order derivative, $\nabla^{2}E_{hi}$, is negligible compared to other terms. This is equivalent to saying that the amplitude envelope function, $E_{h}$ , varies very slowly over a length scale equal to the wavelength or zone width, i.e., only the phase term, $P_{h}$, contains quickly varying components. For this approximation to be valid, one needs to ensure that no specularly reflected wave from the MLL/vacuum boundary is strongly excited. This is because if the incidence angle (the angle between the incidence direction and the MLL surface), is much larger than the critical angle at which total external reflections occurs and the specularly reflected wave is negligible, the approximate solution obtained without second order derivatives can be uniquely determined by the boundary condition for the electric field alone. Under this condition, it is easy to verify that to a good approximation this solution can also satisfy the boundary condition for the magnetic field. In other words, without second order derivative terms, the system is still self-consistent. On the other hand, when the incidence angle approaches the critical angle and the specularly reflected wave is strongly excited, this approximate solution can no longer satisfy the boundary condition for the magnetic field, violating the physical law for electromagnetic field. To have a solution that is able to satisfy the boundary conditions for both electric and magnetic fields, second order derivatives should be retained in the equation. For hard x-rays the critical angle for most materials is smaller than $1^{\circ}$. For example, at 20 keV the critical angle of total external reflection is $0.23^{\circ}$ for tungsten. The specular wave decreases below $10^{-3}$ above $0.67^{\circ}$. If we consider values of the numerical aperture of 0.7 or smaller here, corresponding to incidence angles of $45^{\circ}$ or larger, then the second order derivatives can be neglected. It is important to point out that this approximation of neglecting second order derivatives also impose a restriction on how rapidly $E^{(a)}_{0}$ can vary on the entrance surface, in order to guarantee that the second order derivative terms are negligible everywhere.   

By neglecting second order derivatives we obtain,
\[
	\sum_{i=x,y,z}{(2\nabla E_{hi}\cdot\nabla P_{h}+E_{hi}\nabla^{2}P_{h}+k^{2}E_{hi}P_{h})\vec{e}_{i}}=-k^{2}P_{h}\sum_{l}{\chi_{h-l}\vec{E}_{l}}.
\]
Taking the dot product on both sides with $\vec{E}_{h}$, one obtains a scalar equation,
\begin{equation}
	\nabla E_{h}^{2}\cdot\nabla P_{h}+E_{h}^{2}\nabla^{2}P_{h}+k^{2}E_{h}^{2}P_{h}=-k^{2}P_{h}\sum_{l}\chi_{h-l}E_{h}E_{l}\cos\vartheta_{hl},
	\label{eq10}
\end{equation}
where $\vartheta_{hl}$ is the angle between the polarization directions of two wave components, $\vec{E}_{h}$ and $\vec{E}_{l}$. Eq. (\ref{eq10}) can be further simplified by canceling one factor of $E_{h}$ to give,
\begin{equation}
	2\nabla E_{h}\cdot\nabla P_{h}+E_{h}\nabla^{2}P_{h}+k^{2}E_{h}P_{h}=-k^{2}P_{h}\sum_{l}\chi_{h-l}E_{l}\cos\vartheta_{hl}.
	\label{eq11}
\end{equation}
Because $P_{h}$ is known [see Eq. (\ref{eq8})], we obtain,
\begin{eqnarray*}
&&\nabla P_{h}=i(\vec{k}_{0}+\nabla\phi_{h})P_{h},\vec{k}_{0}=k\vec{s}_{0},\\
&&\nabla^{2}P_{h}={i\nabla^{2}\phi_{h}-[k^{2}+2\vec{k}_{0}\cdot\nabla\phi_{h}+(\nabla\phi_{h})^{2}]}P_{h}.
\end{eqnarray*}
Substituting these expressions into Eq. (\ref{eq11}) yields,
\begin{equation}
	\frac{2i}{k}\nabla E_{h}\cdot(\vec{s}_{0}+\frac{\nabla\phi_{h}}{k})+\beta_{h}(\vec{r})E_{h}+\sum_{l}\chi_{h-l}E_{l}\cos\vartheta_{hl}=0,\ h,l=0,\pm 1,\pm 2, \pm 3,\ldots,
	\label{eq12}
\end{equation}
where
\[
\beta_{h}=i\frac{\nabla^{2}\phi_{h}}{k^{2}}-2\vec{s}_{0}\cdot\frac{\nabla\phi_{h}}{k}-(\frac{\nabla\phi_{h}}{k})
^{2}\approx\frac{k^{2}-(\vec{k}_{0}+\nabla\phi_{h})^{2}}{k^{2}}.
\]
These are the central equations that will be using in this paper. Here $\beta_{h}$ is the deviation function, which quantifies the violation of the Bragg condition. Because an infinite number of differential equations need to be solved simultaneously (both $h$ and $l$ run from -$\infty$ to $\infty$), a system described by Eq. (\ref{eq12}) is considered presently to be still too complicated. By noting that a diffracted wave is strongly excited only near the Bragg condition, we can truncate the system to a finite number of equations. Consequently, it is sufficient to consider a finite number (N) of coupled first order partial differential equations, with the boundary conditions of the electric field at the entrance surface (Laue case)given by,
\begin{equation}
	\vec{E}_{h,h\neq 0}=0,\qquad 
	\vec{E}_{0}=\vec{E}^{(a)}_{0}.
\end{equation}

One may recognize that Eq. (\ref{eq12}) resembles the Takagi-Taupin equations for strained single crystals. This is due to the fact that the methodologies of these two theories are the same: both utilize the prior knowledge about the solution for diffraction from a periodic structure, thereby allowing the separation of the fast and slowly varying components. However, there is a fundamental difference between them: the lattice constant of a strained single crystal differs from its unstrained value only very slightly, and as a result the diffracted wave vector can be regarded approximately invariant with position. In other words, in a strained crystal, the following approximation is valid:
\begin{equation}	
	\vec{s}_{0}+\frac{\nabla\phi_{h}}{k}=\vec{s}_{0}+\frac{\vec{\rho}_{h}}{k}\approx\vec{s}_{0}+\frac{\vec{\rho}_{h_{0}}}{k}\approx\vec{s}_{h},
	\label{eq14}
\end{equation}
where $\vec{\rho}_{h_{0}}$ is the unstrained reciprocal lattice vector and $\vec{s}_{h}$ is a unit vector along the diffracted wave direction. In addition, in a crystal often only one diffracted wave is strongly excited, so only two waves need to be considered. Under these approximations, Eq. (\ref{eq12}) becomes,
\begin{eqnarray}
	\left\{%
	\begin{array}{l}\frac{\partial}{\partial s_{0}}E_{0}=i\frac{\pi}{\lambda}(\chi_{0}E_{0}+C\chi_{\bar{h}}E_{h})\\
	\frac{\partial}{\partial s_{h}}E_{h}=i\frac{\pi}{\lambda}[(\beta_{h}+\chi_{0})E_{h}+C\chi_{h}E_{0}]\end{array}\right.,
	\ \text{where } C=\cos\vartheta_{h0},
	\label{eq15}
\end{eqnarray}
which are the well-known Takagi-Taupin equations for crystal diffraction. For volume diffractive optics its $d$-spacing can vary from hundreds of nanometers in the vicinity of the center to several nanometers or even smaller in the outmost region, so the variation of the diffracted wave vector is significant and the approximation in Eq. (\ref{eq14}) is invalid. Also, for areas of a diffractive optics with large $d$-spacing, many diffraction orders can be excited, and consideration of only two orders is insufficient for the description of volume diffractive optics in these areas. 

\section{DIFFRACTION FROM an MLL WITH FLAT, TILTED and WEDGED ZONES}
\subsection{Flat and Tilted Zones}
\begin{figure}[tbp]
	\centering
		\includegraphics[width=3.2in]{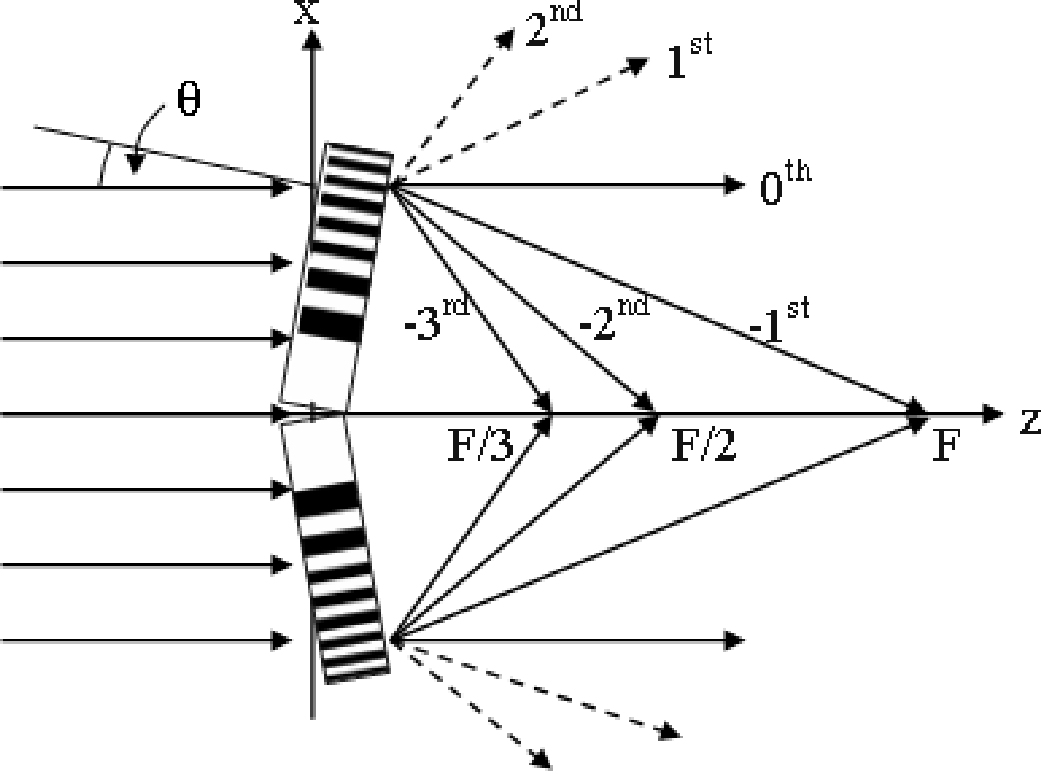}
	\caption{Schematic of a plane wave diffracted by a tilted MLL with flat zones, in which many diffraction orders are excited.}
	\label{fig:FIG4}
\end{figure}

\begin{figure}[tbp]

	\subfigure 
{
    \label{fig:FIG5:a}
    \includegraphics[width=2.5in]{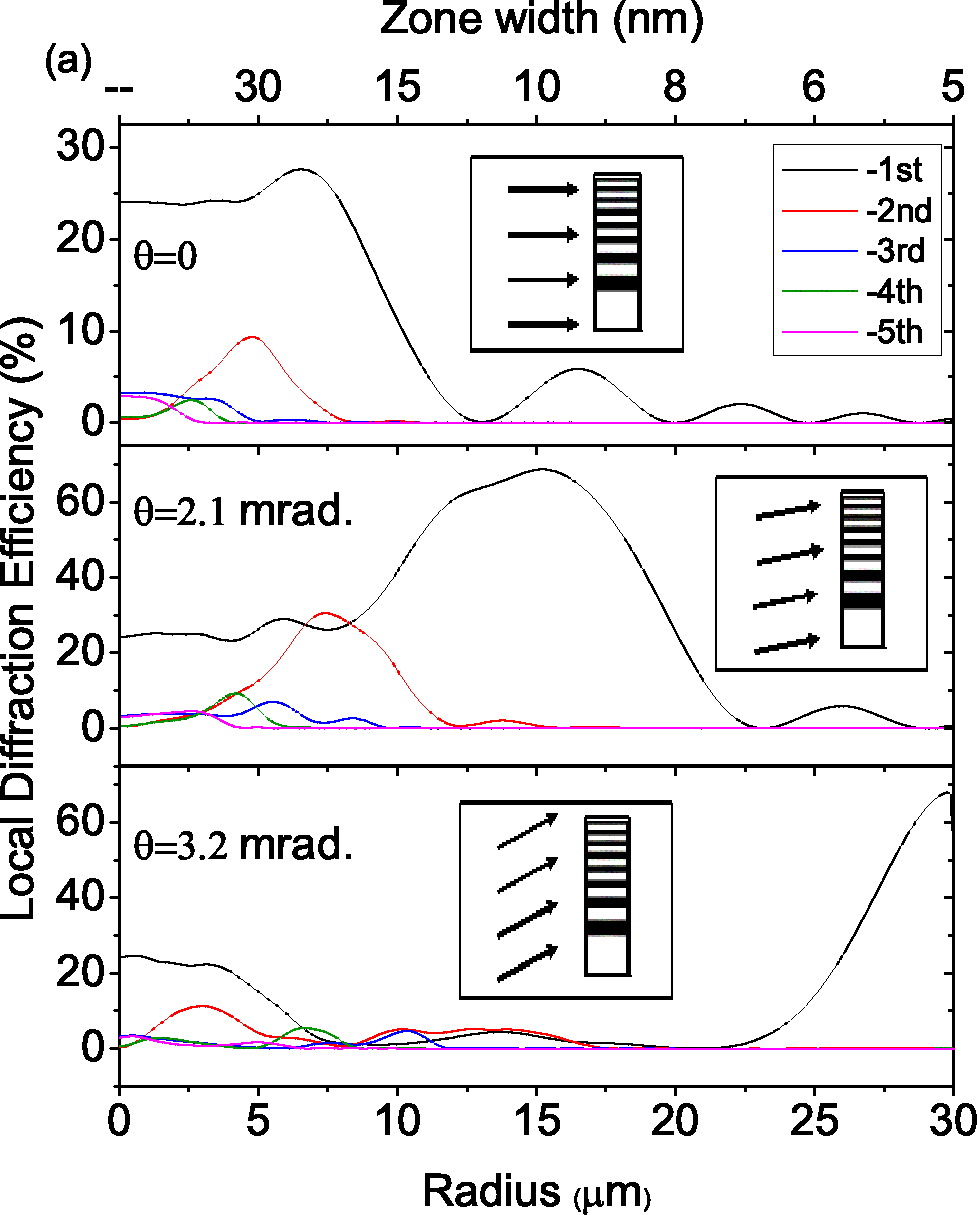}
}
	\subfigure 
{
    \label{fig:FIG5:b}
    \includegraphics[width=3.2in]{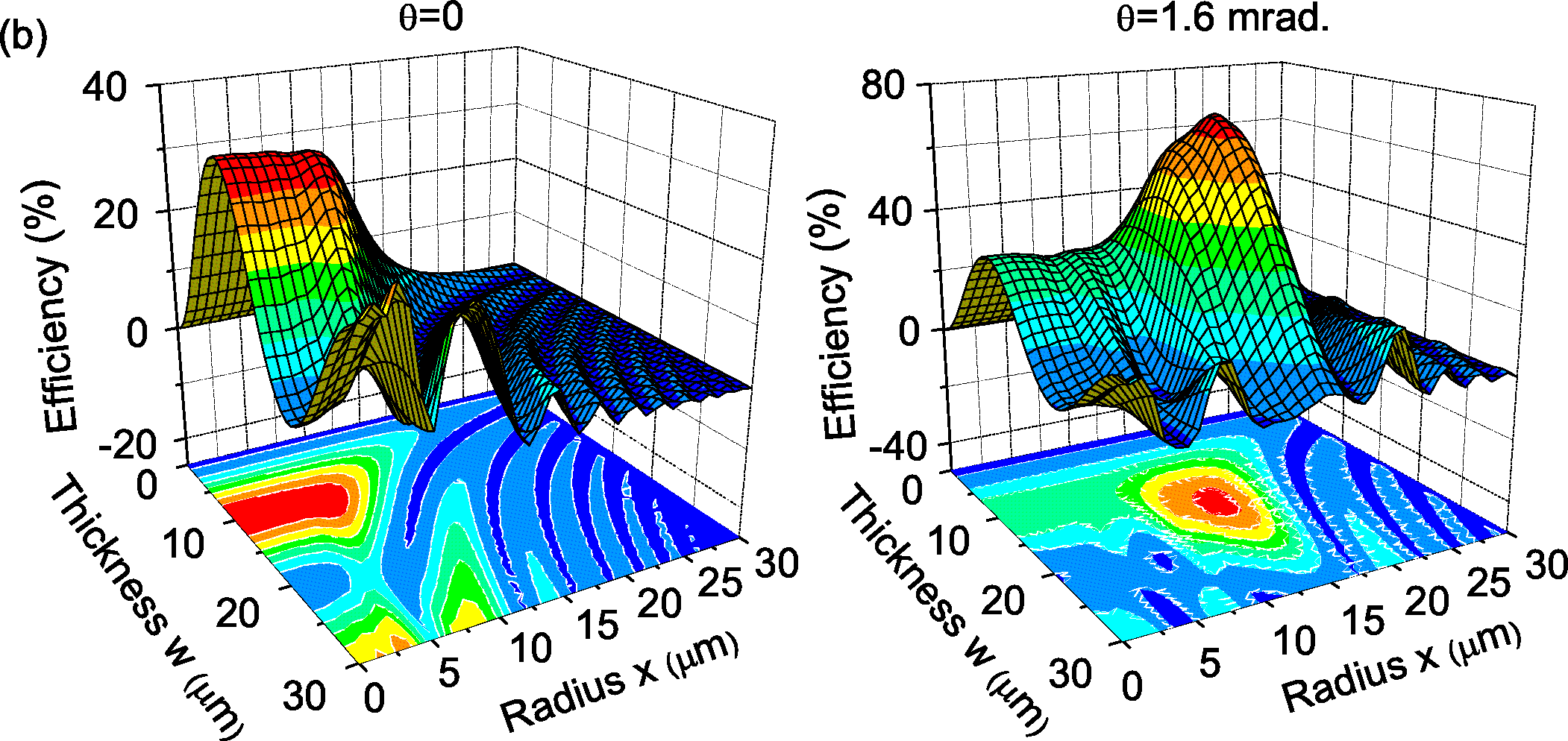}
}
\subfigure
{
    \label{fig:FIG5:c}
    \includegraphics[width=2.5in]{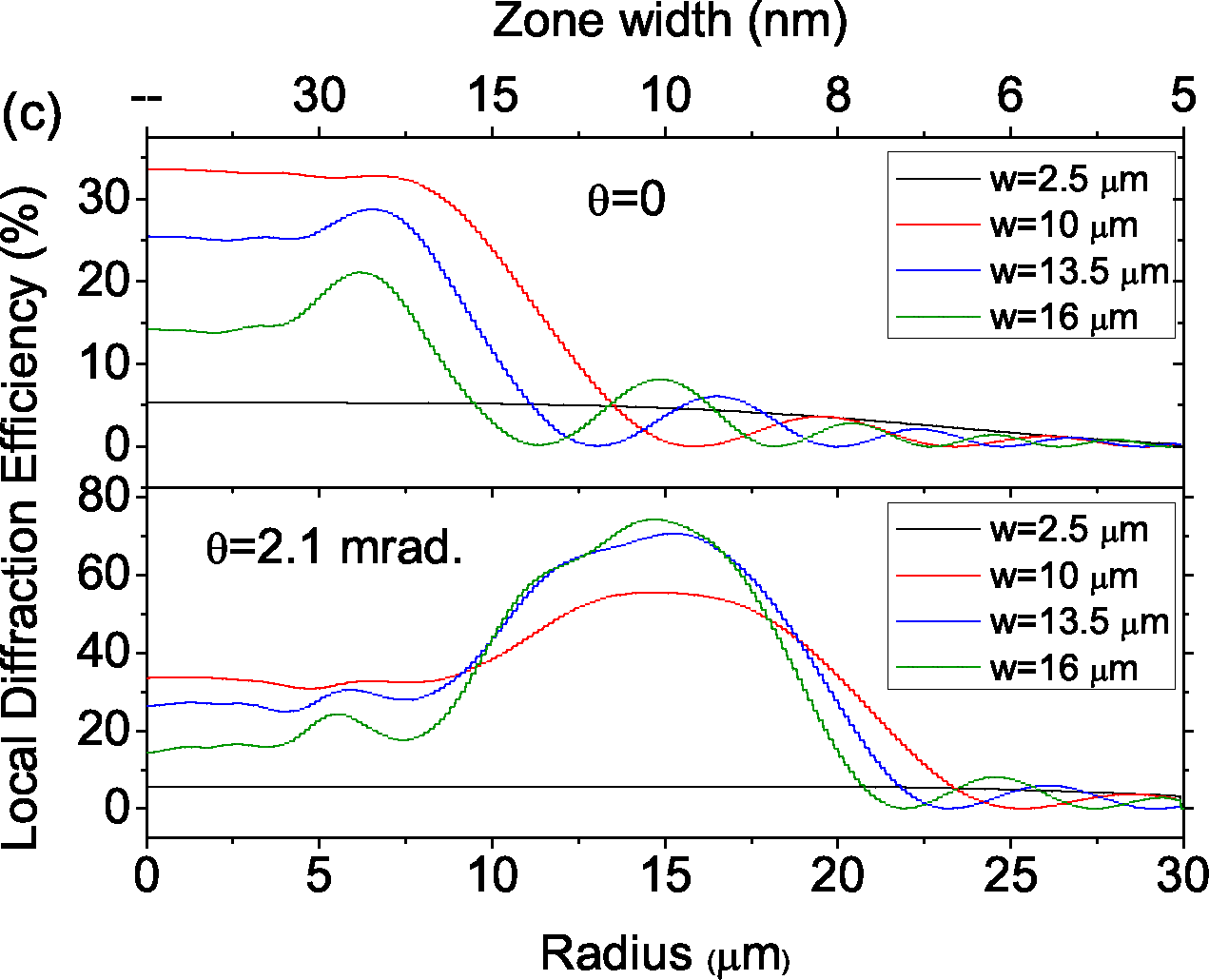}	
}
	\caption{a) The radial efficiency distributions of different diffraction orders on the exit surface of a flat MLL at $\theta=0$ (top), $\theta=1.6$ (middle) and $\theta=3.2$ mradian (bottom).  b) The efficiency variation of the $-1^{st}$ order diffraction inside the zone plate at $\theta=0$ (left) and $\theta=1.6$ mradian (right). c) The radial efficiency distribution of the $-1^{st}$ order diffraction on the exit surface at different thickness and tilting angles.}
	\label{fig:FIG5}
\end{figure}

Using a Takagi-Taupin description for volume diffractive optics, we can study dynamical diffraction from various types of MLL's. We will discuss flat, tilted and wedged MLL's. Flat and tilted MLL's have zones that are parallel to each other, but are aligned at different tilt angle with respect to the incident beam. Wedged MLL's have zones that are tilted with respect to each other, such as to satisfy the Bragg angle with regard to the direction of the incident wave. We consider MLL's with a diameter of $30\ \mu$m, which corresponds to typical lateral coherence lengths of 20 keV x-rays at a $3^{rd}$ generation synchrotron source, and a thickness of $13.5\ \mu$m, which corresponds to the optimum thickness of an MLL of Si and WSi2 for a photon energy of 19.5 keV. As an example, we use an outermost zone width of 5 nm, corresponding to MLL structures we fabricated currently\cite{KangUnpublished}. A plane wave at 19.5 keV is impinging on the MLL described above, with an inclination angle of $\theta$ to the normal direction of the sample surface, as shown in Fig. \ref{fig:FIG4}. For simplicity, $\sigma$-polarization, corresponding to $\cos\vartheta_{hl}=1$, is assumed. In this case $x_{max}=30\ \mu$m is much smaller than the focal length $f=4.72$ mm and the approximate zone plate law can be used, so,
\[
x_{n}=\sqrt{n\lambda f},\ \text{and $\phi_{h}=\frac{h\pi}{\lambda f}x^{2}$.}
\]
From Eq. (\ref{eq12}) one can deduce,
\begin{eqnarray*}
&&\beta_{h}(x)\approx-2\frac{hx}{f}\sin\theta -(\frac{hx}{f})^2,\\
&&\frac{\nabla\phi_{h}}{k}=\frac{hx}{f}\vec{e}_{x}.
\end{eqnarray*}
In order to obtain a complete description of the wavefield, 11 beams ($h=0, \pm 1, \pm 2, \pm 3, \pm 4, \pm 5$) are considered. Thus, we need to obtain a solution for eleven coupled hyperbolic partial differential equations given by,
\begin{equation}
	[(\sin\theta+\frac{h}{f}x)\frac{\partial}{\partial x}+\cos\theta\frac{\partial}{\partial z}]E_{h}=i\frac{\pi}{\lambda}(\sum^{5}_{l=-5}\chi_{h-l}E_{l}+\beta_{h}E_{h}),\qquad h=-5,-4,...4,5,
	\label{eq16}
\end{equation}
with the boundary conditions, $E_{h,h\neq 0}(x,0)=0$ and $E_{0}(x,0)=1$. Eq. (\ref{eq16}) is solved numerically in Mathematica 5.2, and the simulation results are displayed in Fig. \ref{fig:FIG5:a} for $\theta$ equal to 0, 1.6 and 3.2 mrad., respectively. Since we are not interested in the diffractions with positive orders which correspond to divergent waves, they are not plotted. We find, as shown in the figures, that higher order diffractions have weaker intensities and the fifth order diffraction intensity becomes negligible. This justifies the approximation of only retaining diffraction orders up to order five. 

For normal incidence ($\theta=0$), the intensity distribution of all diffracted beams show a similar trend: only in the vicinity of the center ($x=0$) do the diffracted beams have appreciable intensities; below a certain $d$-spacing the diffracted intensity decreases to zero quickly. The local diffraction efficiency, defined as the ratio of the local diffraction intensity to the incident wave intensity, which is unity, is below 30\% for all diffraction orders. Near the center where the value of the deviation function, $\beta_{h}$, is relatively small, a large number of diffracted waves are excited. The value of the deviation function increases rapidly as the $d$-spacing becomes smaller with increasing radius, causing a fast drop of the diffraction intensity in the outmost region. This indicates that only a fraction of the MLL near the center contributes to the focusing. This is consistent with the results obtained by Maser\cite{Maser1994} and by Pfeiffer et al.\cite{Pfeiffer2006}. The radial variation of the diffraction efficiency also depends on the thickness of the MLL, as seen in Fig. \ref{fig:FIG5:b} where the $-1^{st}$ order diffraction intensity profile is plotted as a function of radius and thickness. When the thickness is very small, the diffraction efficiency is low, and no dynamical effects are encountered. At larger thickness, the efficiency increases for areas with larger $d$-spacing, but stays small for areas with smaller $d$-spacing, where deviations from the Bragg condition have an appreciable effect. This trend is clearly shown in Fig. \ref{fig:FIG5:c}, where the radial efficiency distribution of the $-1^{st}$ order at different thickness is plotted. A maximum diffraction efficiency, about 34\%, is reached at $w=10\ \mu$m. This number agrees with the value of optimum thickness for a phase zone plate obtained in geometrical optics, $w=\lambda/\Delta\chi\approx 10\ \mu$m\cite{Kirtz1974}.

When the MLL is tilted, the Bragg condition can be satisfied for a small area of the MLL. The middle graph in Fig. \ref{fig:FIG5:a} shows the local diffraction efficiency at a tilt angle of $\theta=1.6$ mrad. At this angle the Bragg condition is satisfied for the $-1^{st}$ diffraction order at $x=15\ \mu$m, corresponding to a zone width of 10 nm. A diffraction efficiency maximum is observed at $x=15\ \mu$m, with a local diffraction efficiency of 67\%, as expected from dynamical diffraction. At this tilting angle the local diffraction efficiency starts to decrease at a larger radius than that in the case of normal incidence, indicating that a larger fraction of the MLL contributes to the focusing. An increase in the diffraction efficiency of the $-2^{nd}$ (focusing) diffraction order is observed for a radius of $7.5\mu$m, corresponding to a zone width of 20 nm, where the Bragg condition is satisfied for the $-2^{nd}$ order. In Fig. \ref{fig:FIG5:b} the variation of the local efficiency of the $-1^{st}$ order diffraction as a function of radius and thickness is displayed. The bottom graph in Fig. \ref{fig:FIG5:c} shows its radial efficiency distribution at different thicknesses. We observe at a thickness of $16\ \mu$m, a maximum diffraction efficiency of the $-1^{st}$ order is reached. In the phenomenon of ``Pendellösung fringes'' for dynamical diffraction from a single crystal in Laue geometry, a Laue peak maximum is reached at a thickness of $w=\lambda \cos\theta/2\sqrt{\left|\chi_{-1}\chi_{1}\right|}\approx16\ \mu$m\cite{Authier2002}, which is in a reasonable agreement with the value obtained here. This agreement suggests that dynamical resonant scatterings are strongly excited in this case, which raises the diffraction efficiency as well as the effective radius.

Here the even order diffractions are excited via dynamical multiwave scattering processes, similar to the Renninger effect in diffraction from crystals\cite{Speakman1965}, although they are forbidden in the geometrical-optical theory\cite{Kirtz1974}.

\begin{figure}[tbp]
	\subfigure 
{
    \includegraphics[width=3.2in]{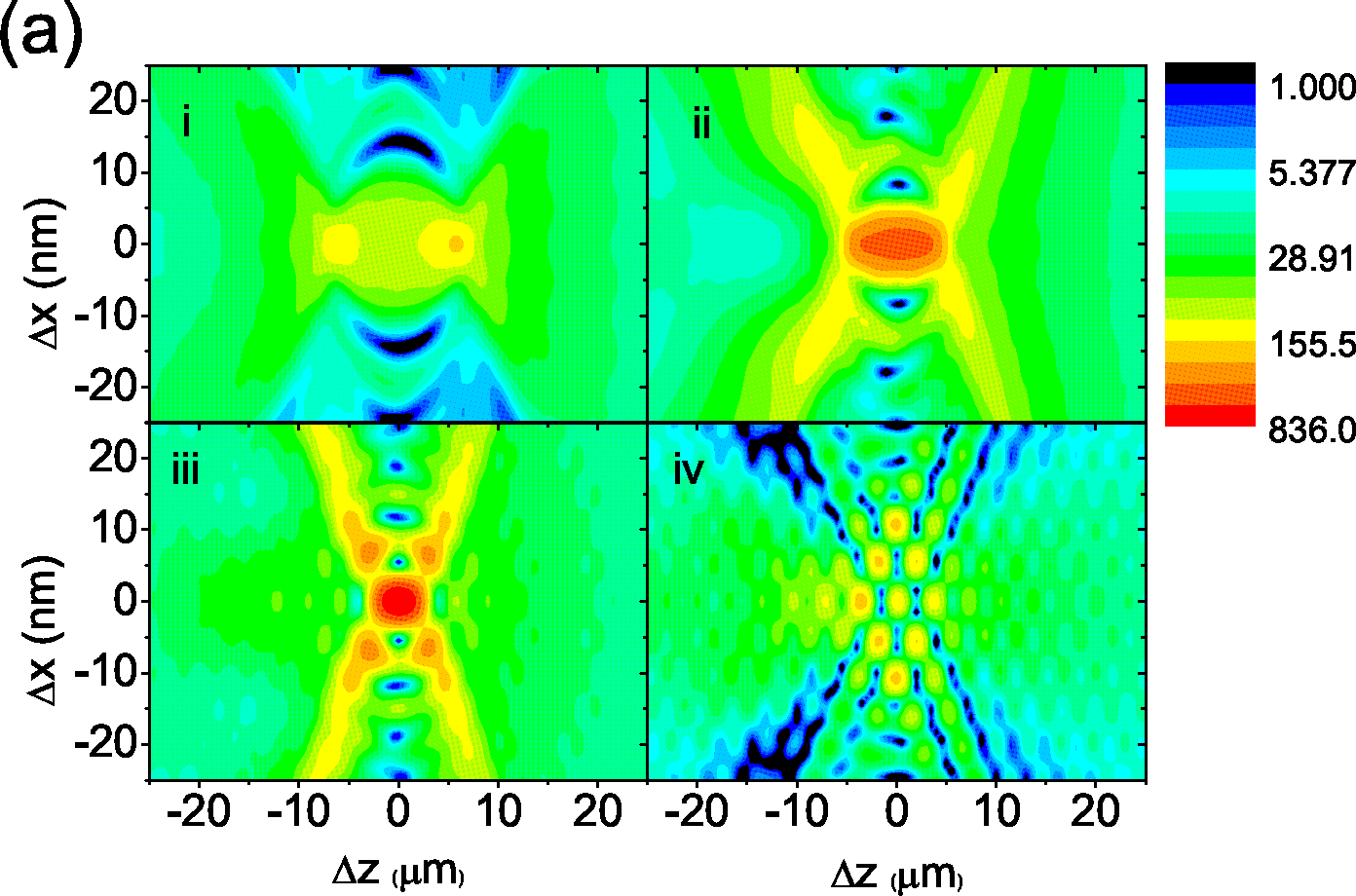}
    \label{fig:FIG6:a}
}
	\subfigure 
{
    \includegraphics[width=3.2in]{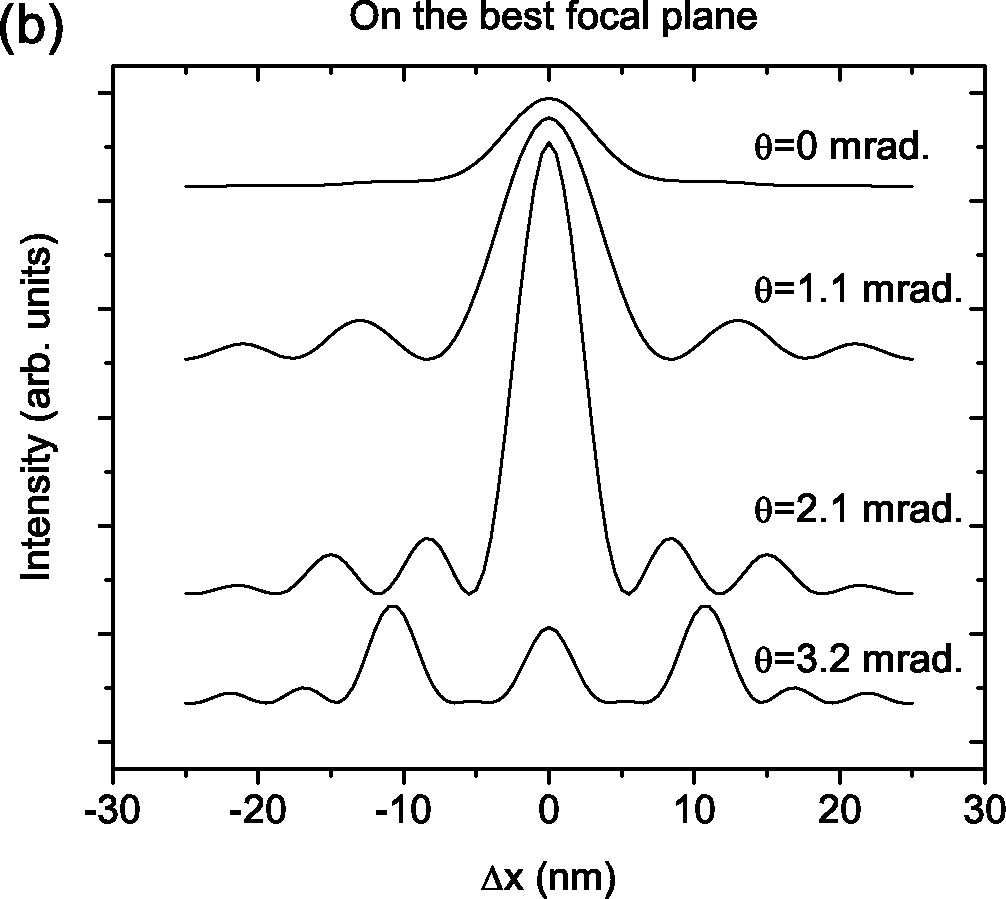}
    \label{fig:FIG6:b}
}	
	\caption{a) Isophotes near the focus at different tilting angles. $\theta=0,\ 1.1,\ 2.1\ \text{and}\ 3.2$ mrad. in i), ii), iii) and iv). The intensity is plotted in logarithm scale. b) The intensity profiles on the best focal plane. }
	\label{fig:FIG6}
\end{figure}

To study the focusing property of the flat and tilted MLL's, we employed Fresnel-Kirchhoff integral\cite{Born1999} to calculate the wavefield at a point behind the MLL. In Fig. \ref{fig:FIG6:a} the isophotes near the focus at different tilting angles are plotted. As expected, we observe in our calculations that tilting leads to a reduction in beam size and to an enhancement of peak intensity. However, there is an optimum tilting angle above which defocusing occurs. This tendency is clearly seen in Fig. \ref{fig:FIG6:b} by comparing the line focus profiles on the best focal plane at different tilting angles. When $\theta=0$, the full width at half maximum (FWHM) of the best focused beam is about 7.4 nm. As $\theta$ increases to 2.1 mradian, the FWHM of the best focused beam is reduced to 4.7 nm because a larger fraction of the MLL contributes to the focusing, as shown in Fig. \ref{fig:FIG5:a}. However, when $\theta=3.2$ mradian at which Bragg condition is satisfied for the outmost zones at $x=30\ \mu$m, multiple peaks with nearly equal intensities appear on the plane of best focus, destroying the resolution. This is a result of the uneven radial intensity distribution at $\theta=3.2$ mradian (see the bottom graph in Fig. \ref{fig:FIG5:a}). At this tilting angle, the diffraction efficiency of the $-1^{st}$ order  has appreciable values only near the center and near the outmost regions where a Bragg condition is satisfied, while in between the diffraction efficiency is very low. As a result, the diffracted waves from the center and the outmost region interfere with each other strongly, causing multiple peaks with nearly equal intensity on the focal plane in an analogy to the multi-slit diffraction pattern. The effective radius within which the diffraction efficiency is nonzero constrains the focal size of a tilted MLL with parameters given here to slightly smaller than 5nm. Therefore, if the radius of the MLL is increased to $60\ \mu$m and the outmost zone width is reduced to 2.5 nm, at this optimum thickness we still cannot obtain a 2.5-nm beam size. To increase the effective radius, we need to reduce the thickness of the MLL, but with a significant loss of efficiency, as shown in Fig. \ref{fig:FIG5:c}. 

The improvement of the focusing by tilting might be better understood by considering the structure of an MLL, which consists of many thin layers. The diffraction property of an MLL is determined by the scattering property of a single layer and the interference effect of the scattered waves from different layers as well. Each thin layer can be regarded as a waveguide where total external reflection occurs at the boundaries. As Bergemann et al. argued\cite{Bergemann2003}, the critical angle of the total external reflection imposes an ultimate limit on the achievable NA for a waveguide, because above this critical angle x-rays start to leak out, resulting in a rapid drop of the reflection intensity. Consequently the maximum converging angle of x-rays from a waveguide can not exceed the critical angle, so the maximum achievable NA for a waveguide is about $\sqrt{2\Delta n}$, where $\Delta n$ is the difference of the refractive index of the waveguide material from unity. They also claimed that this limit applies to all x-ray focusing optics. However, it has been shown that this limit could be overcome by a thick adiabatic refractive lens\cite{Schroer2005a}. Such a limit does not apply to diffractive optics either since resonant scattering occurs when the Bragg condition is fulfilled, so that the reflected waves from different layers can interfere constructively, leading to a resultant wave that can be as strong as the incident wave. For the MLL the converging angle is only limited by the Bragg angle, which can be as large as $\pi/2$. Though the waveguide effect will somewhat influence the achievable NA of an MLL when Bragg condition is not satisfied ($\theta=0$), it does not determine the ultimate resolution of an MLL.    

\subsection{Wedged MLL}
As shown previously, tilted MLL's can focus x-rays much more effectively than flat MLL's. However, tilted MLL cannot achieve a 1-nm focusing without significantly sacrificing the efficiency. To achieve both high efficiency and small focusing, we consider a structure in which there is a distinct tilt to every zone by an angle increasing with radius so that the local Bragg condition is fulfilled in every zone. This geometry corresponds to a wedged zone shown in Fig. \ref{fig:FIG1}. Deducing from Eqs. (\ref{eq2}) and (\ref{eq3}), we obtain the modified zone plate law corresponding to an MLL with wedged zones which satisfy the Bragg condition at $z=0$,
\begin{equation}
	x_{n}=a_{n}(z)\sqrt{n\lambda f+\frac{n^{2}\lambda^{2}}{4}},\qquad a_{n}=1-\frac{z}{2f(1+n\lambda/4f)}.
	\label{eq17}
\end{equation}
At the entrance surface ($z=0$) the zone width obeys the conventional zone plate law [Eq. (\ref{eq2})], but all zones shrink by a factor $a_{n}(z)$ along the depth $z$. The shrinkage factor, $a_{n}$, also depends on the zone number $n$. To simplify the calculation, as an approximation we consider a shrinkage factor independent of $n$ given by,   $a_{n}=1-z/[2f(1+n_{max}\lambda/8f)]$, where $n_{max}$ is the maximum zone number. Consequently all zones shrink homogeneously by a factor proportional to the depth $z$,
\[
\Delta x_{n}=a(z)\frac{\lambda f}{2x}\sqrt{1+\frac{x^{2}}{f^{2}}},
\]
and the phase function $\phi_{h}$ becomes,
\begin{equation}
	\phi_{h}=hk[\sqrt{\frac{x^{2}}{a(z)^{2}}+f^{2}}-f],\qquad h=0,\pm 1, \pm 2, \pm 3,\ldots
	\label{eq18}
\end{equation}

\begin{figure}[tbp]
	\subfigure 
{
    \label{fig:FIG7:a}
    \includegraphics[width=3.2in]{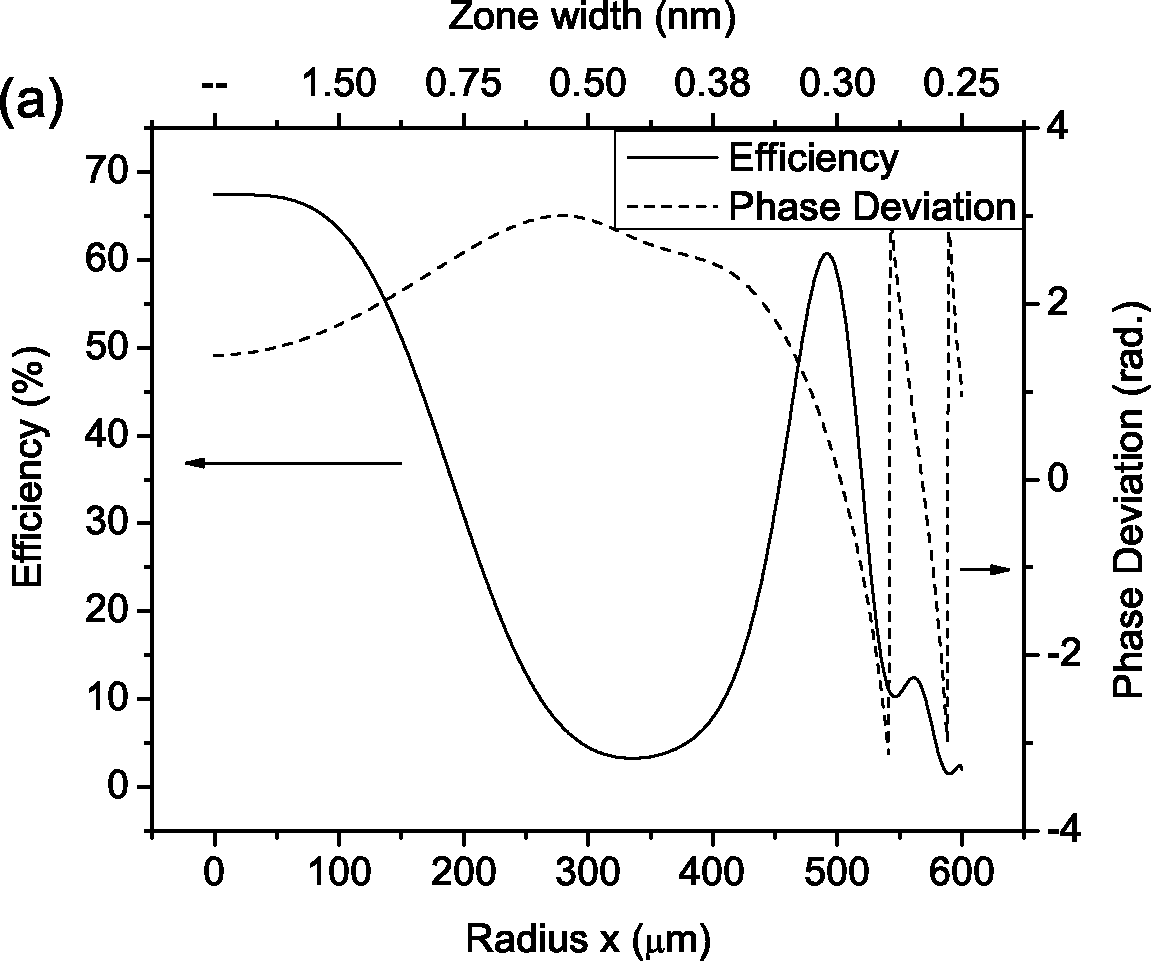}
}
	\subfigure 
{
    \label{fig:FIG7:b}
    \includegraphics[width=3.2in]{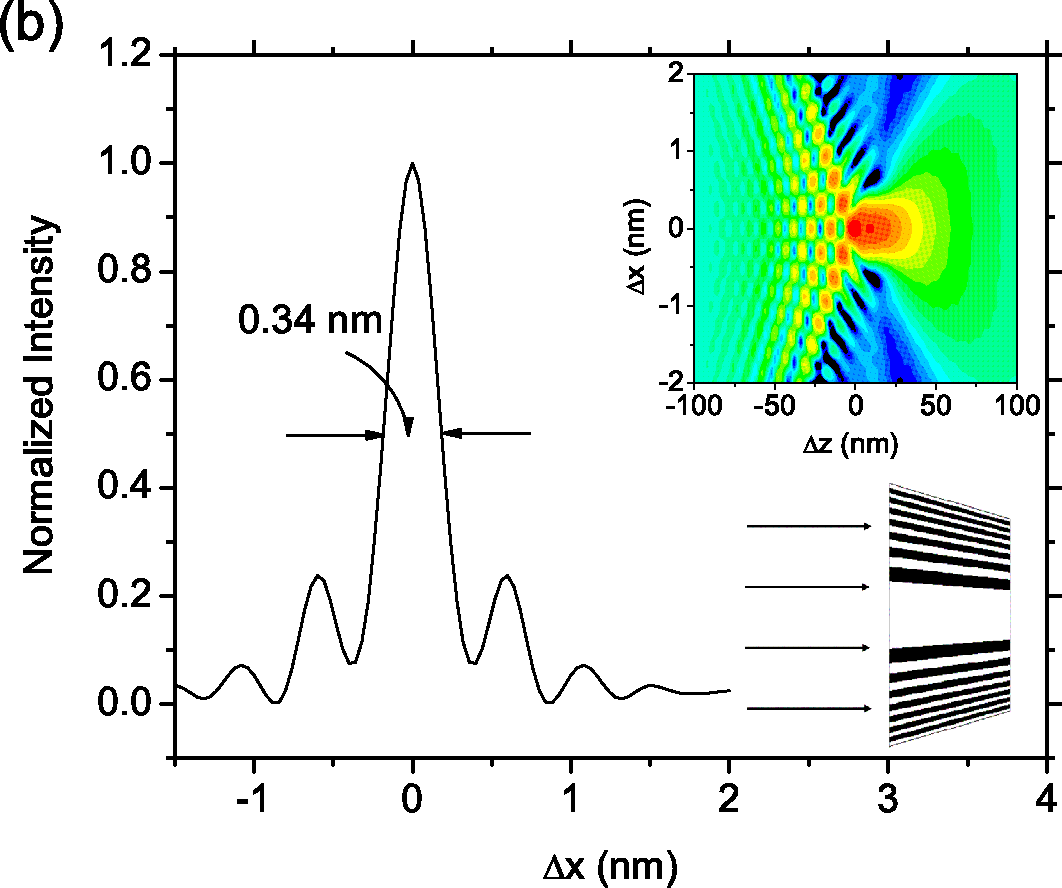}
}
	\caption{a) The radial intensity distribution of the -1st order diffraction on the exit surface of a wedged MLL with an outmost zone width of 0.25 nm. Two-beam approximation is assumed. The dashed curve represents the phase deviation from a perfect converging spherical wave front. b) The intensity profile on the focal plane, showing a peak width FWHM=0.34 nm. The inset on top is the isophotes near focus, intensity in logarithm scale. The one on bottom is a sketch of the wedged MLL structure.}
	\label{fig:FIG7}
\end{figure}

To explore a possible upper limit of the achievable NA, we consider an MLL with following parameters: $f=4.72$ mm, $w=13.5\ \mu$m, $E=19.5$ keV, $x_{max}=600\ \mu$m and an outmost zone width of 0.25 nm. As we demonstrated previously [see Fig. \ref{fig:FIG5:a}], for zone widths less than 10 nm high diffraction orders are negligible and the two-beam approximation is sufficient to describe the diffraction property of an MLL. Since here we are interested in the diffraction behavior of zones with much smaller widths, from now on, unless otherwise specified, the two-beam approximation will be used. Simulation results are shown in Fig. \ref{fig:FIG7:a} and \ref{fig:FIG7:b}. In Fig. \ref{fig:FIG7:a} the radial efficiency distribution and the phase deviation of the diffracted wave from a perfect spherical wave that converges to the focal point are plotted. Although in the outmost region the diffracted waves still have strong intensities, they do not contribute to the focus because their phase deviation varies rapidly over $\pi$ and they interfere destructively at the focal spot. Therefore, using this structure, we cannot focus an incident plane wave down to the diffraction limit due to the phase effect. In Fig. \ref{fig:FIG7:b} the intensity profile on the best focal plane is displayed. The inset on top shows the isophotes near the focus, and the bottom one is a sketch of the zone plate structure. In this case the achievable NA is also limited by the dynamical diffraction property of the MLL and a larger physical radius will not increase its NA. This study indicates that to effectively focus x-rays down to the wavelength using diffractive optics, not only the Bragg condition has to be satisfied so that the diffracted wave intensity is strong and the efficient is high, but also the phase has to be right so that all diffracted waves add up in phase at the focal spot. We will derive the optimum curved shapes that minimize the radial phase change in section IV. 

\subsection{Localized One-Dimensional Theory}
\begin{figure}[tbp]
	\centering
		\includegraphics[width=3.2in]{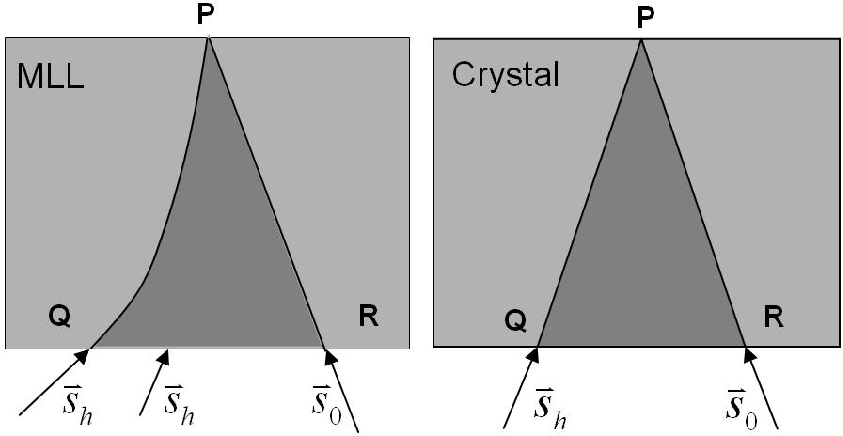}
	\caption{The domain of dependence in an MLL (left), which has a curved edge due to the change of diffracted direction with positions, and the inverse Borrmann fan in a single crystal (right), which is a triangle that is invariant with position.}
	\label{fig:FIG8}
\end{figure}

In this section we show that under certain conditions the assumption of local perfect gratings becomes valid and the basic diffraction equations derived in this paper are simplified to simpler forms similar to those in 1-D CWT. For the sake of simplicity, we start with the two-beam approximation, so that Eq. (\ref{eq16}) is reduced to,
\begin{equation}
\left\{\begin{array}{l}(\sin\theta\frac{\partial}{\partial x}+\cos\theta\frac{\partial}{\partial z})E_{0}=i\frac{\pi}{\lambda}(\chi_{0}E_{0}+\chi_{\bar{h}}E_{h})\\ \lbrack(\sin\theta+\frac{hx}{f})\frac{\partial}{\partial x}+\cos\theta\frac{\partial}{\partial z}\rbrack E_{h}=i\frac{\pi}{\lambda}[(\chi_{0}+\beta_{h})E_{h}+\chi_{h}E_{0}]\end{array}\right.,\qquad h=\pm 1, \pm 3, \pm 5,\ldots
	\label{eq19}
\end{equation}
To simplify the calculation, dimensionless variables are used,
\[
\bar{x}=\frac{hx}{f \sin\theta},\qquad \bar{z}=\frac{hz}{f \cos\theta}.
\]
Consequently we obtain,
\begin{equation}
	\left\{\begin{array}{ll}(\frac{\partial}{\partial \bar{x}}+\frac{\partial}{\partial \bar{z}})E_{0}=i\frac{f\pi}{h\lambda}(\chi_{0}E_{0}+\chi_{\bar{h}}E_{h})\\\lbrack(1+\bar{x})\frac{\partial}{\partial \bar{x}}+\frac{\partial}{\partial \bar{z}}\rbrack E_{h}=i\frac{f\pi}{h\lambda}[(\chi_{0}+\beta_{h})E_{h}+\chi_{h}E_{0}]\end{array}\right.,
	\label{eq20}
\end{equation}
where $\beta_{h}\approx-\bar{x}(2+\bar{x})\sin^{2}\theta$. We may further simplify Eq. (\ref{eq20}) by setting,
\begin{align*}
E_{0}=\widetilde{E}_{0}\exp(i\frac{f\pi}{h\lambda}\chi_{0}\bar{z}),\\
E_{h}=\widetilde{E}_{h}\exp(i\frac{f\pi}{h\lambda}\chi_{0}\bar{z}),
\end{align*}
so the refraction and photoelectric absorption effects are taken out of the equation. Substituting above expressions into Eq. (\ref{eq20}) yields,
\begin{equation}
	\left\{\begin{array}{ll}(\frac{\partial}{\partial \bar{x}}+\frac{\partial}{\partial \bar{z}})\widetilde{E}_{0}=i\frac{f\pi}{h\lambda}\chi_{\bar{h}}\widetilde{E}_{h}\\\lbrack(1+\bar{x})\frac{\partial}{\partial \bar{x}}+\frac{\partial}{\partial \bar{z}}\rbrack \widetilde{E}_{h}=i\frac{f\pi}{h\lambda}(\beta_{h}\widetilde{E}_{h}+\chi_{h}\widetilde{E}_{0})\end{array}\right.,
	\label{eq21}
\end{equation}
which are two coupled hyperbolic partial differential equations. According to the partial differential equation theory\cite{Copson1975}, a solution at a point ($\bar{x}_{0},\bar{z}_{0}$) to a hyperbolic equation only depends on the initial data in a bounded domain of dependence; the outside part of this domain has no influence on the value of the solution at ($\bar{x}_{0},\bar{z}_{0}$). For a system described by Eq. (\ref{eq21}), the domain of dependence is bounded by its characteristics,
\begin{equation}
\bar{x}=(\bar{x}_{0}+1)\exp(\bar{z}-\bar{z}_{0})-1,\qquad \bar{x}=\bar{x}_{0}+(\bar{z}-\bar{z}_{0}).
	\label{eq22}
\end{equation}
In Fig. \ref{fig:FIG8} the domain of dependence of point \textbf{P} is depicted. The counterpart concept in crystallography is the inverse Borrmann triangle in a single crystal, where the wavefield at \textbf{P} is determined by the wavefields in the triangle \textbf{RPQ} (Fig. \ref{fig:FIG8}). A physical elucidation of this concept is that because waves propagate along either $\vec{s}_{0}$ or $\vec{s}_{h}$, the wavefield at \textbf{P} can only be influenced by the wavefields at points that can propagate to \textbf{P}. Such points form a domain of dependence bounded by \textbf{RPQ}. In single crystals both $\vec{s}_{0}$ and $\vec{s}_{h}$ are invariant, so the shape of the inverse Borrmann triangle is independent of position. However in an MLL the diffracted beam direction $\vec{s}_{h}$ is a function of position because the change of reciprocal lattice vector causes a curved edge \textbf{QP} which will vary with position. 

By noting that for $w<<f$, $\bar{z}<<1$, one can make the following approximation,
\[
\exp(\bar{z}-\bar{z}_{0})\approx 1+\bar{z}-\bar{z}_{0}.
\]
Then the first equation in Eq. (\ref{eq22}) becomes,
\[
\bar{x}=\bar{x}_{0}+(\bar{x}_{0}+1)(\bar{z}-\bar{z}_{0}).
\]
The physical manifestation of this approximation is to replace the curved edge \textbf{QP} by a straight line, so Eq. (\ref{eq21}) can be simplified. With the replacement of variables,
\begin{eqnarray*}
&&\xi=\bar{z}-\bar{x},\\ &&\eta=\bar{x}-(\bar{x}+1)\bar{z},
\end{eqnarray*}
one obtains,
\begin{eqnarray*}
&&\frac{\partial}{\partial \bar{x}}+\frac{\partial}{\partial \bar{z}}=-(\bar{x}+\bar{z})\frac{\partial}{\partial \eta},\\
&&(1+\bar{x})\frac{\partial}{\partial \bar{x}}+\frac{\partial}{\partial \bar{z}}=-\bar{x}\frac{\partial}{\partial \xi}-(1+\bar{x})\bar{z}\frac{\partial}{\partial \eta}.
\end{eqnarray*}
In case $\theta$ is very small so that $\bar{z}<<\bar{x}$ can be satisfied, the following approximations are valid,
\begin{eqnarray*}
&&\frac{\partial}{\partial \bar{x}}+\frac{\partial}{\partial \bar{z}}\approx-\bar{x}\frac{\partial}{\partial \eta},\\
&&(1+\bar{x})\frac{\partial}{\partial \bar{x}}+\frac{\partial}{\partial \bar{z}}\approx-\bar{x}\frac{\partial}{\partial \xi}.
\end{eqnarray*}
Use of these approximations leads to,
\begin{equation}	
\left\{\begin{array}{ll}\frac{\partial}{\partial\eta}\widetilde{E}_{0}=-i\frac{f\pi}{h\lambda}\frac{\chi_{\bar{h}}}{\bar{x}(\xi,\eta)}\widetilde{E}_{h}(\xi,\eta)\\ \frac{\partial}{\partial\xi}\widetilde{E}_{h}=-i\frac{f\pi}{h\lambda}[\frac{\beta_{h}(\xi,\eta)}{\bar{x}(\xi,\eta)}\widetilde{E}_{h}(\xi,\eta)+\frac{\chi_{h}}{\bar{x}(\xi,\eta)}\widetilde{E}_{0}(\xi,\eta)]\end{array}\right..
\label{eq23}
\end{equation}
If we note the fact that the wavefield at \textbf{P} is only affected by the wavefields inside the triangle \textbf{RPQ}, and under conditions $\bar{z}<<\bar{x}$ and $\bar{z}<<1$ the value of $\bar{x}$ and $\beta_{h}$ change only very slightly inside this triangle, it is justified to replace variables $\bar{x}$ and $\beta_{h}$  by their values at \textbf{P}. In this way we arrive at two equations that can be solved analytically:
\begin{equation}	
\left\{\begin{array}{ll}\frac{\partial}{\partial\eta}\widetilde{E}_{0}=i\widetilde{\chi}_{\bar{h}}\widetilde{E}_{h}(\xi,\eta)\\ \frac{\partial}{\partial\xi}\widetilde{E}_{h}=i\widetilde{\beta}_{h}(\xi,\eta)\widetilde{E}_{h}(\xi,\eta)+i\widetilde{\chi}_{h}\widetilde{E}_{0}(\xi,\eta)\end{array}\right.,
\label{eq24}
\end{equation}
where,
\begin{eqnarray*}
&&\widetilde{\chi}_{h,\bar{h}}=-\frac{f\pi}{h\lambda}\frac{\chi_{h,\bar{h}}}{\bar{x}(P)},\\
&&\widetilde{\beta}_{h}=-\frac{f\pi}{h\lambda}\frac{\beta_{h}(P)}{\bar{x}(P)},
\end{eqnarray*} are constants. One may recognize that Eq. (\ref{eq24}) coincides with Takagi-Taupin equations for perfect single crystals, implying our approximations have the following physical meaning: locally an MLL can be treated as a strictly periodic grating. But we want to emphasize that the validity of these approximations is limited to the case where $\theta$ is small (corresponding to a small NA), and $w<<f$, i.e., within the triangle \textbf{RPQ} the variation of periodicity can be neglected and MLL can be considered as a periodic grating locally. We found that there were similar discussions about the validity of local 1-D CWT in volume holography previously\cite{Syms1982a,Syms1981}. If all beams are retained in the equation, the discussion is the same but Eq. (\ref{eq24}) will become N-beam Takagi-Taupin equations for perfect single crystals. From Eq. (\ref{eq24}), a close-form analytical solution can be obtained.
 
\section{IDEAL STRUCTURES FOR HIGH NUMERICAL APERTURE}
\begin{figure}[tbp]
	\centering
		\subfigure
		{
		\label{fig:FIG9:a}
		\includegraphics[width=3.2in]{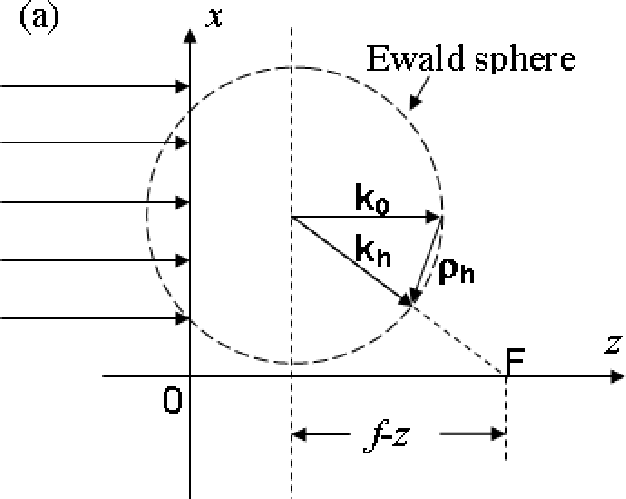}
		}
			\subfigure
		{
		\label{fig:FIG9:b}
		\includegraphics[width=3.2in]{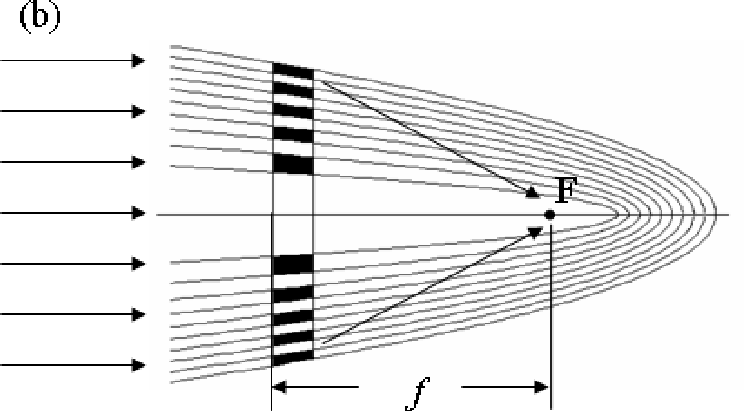}
		}
	\caption{a) A schematic of satisfying Bragg's law everywhere for an incident plane wave. All diffracted wave vectors point to the focal point F. b) A structure consisting of a set of confocal paraboloids obtained from a), serving as an ideal structure to focus an plane wave into a point.}
	\label{fig:FIG9}
\end{figure}

\begin{figure}[tbp]
	\centering
		\subfigure
		{
		\label{fig:FIG10:a}
		\includegraphics[width=3.2in]{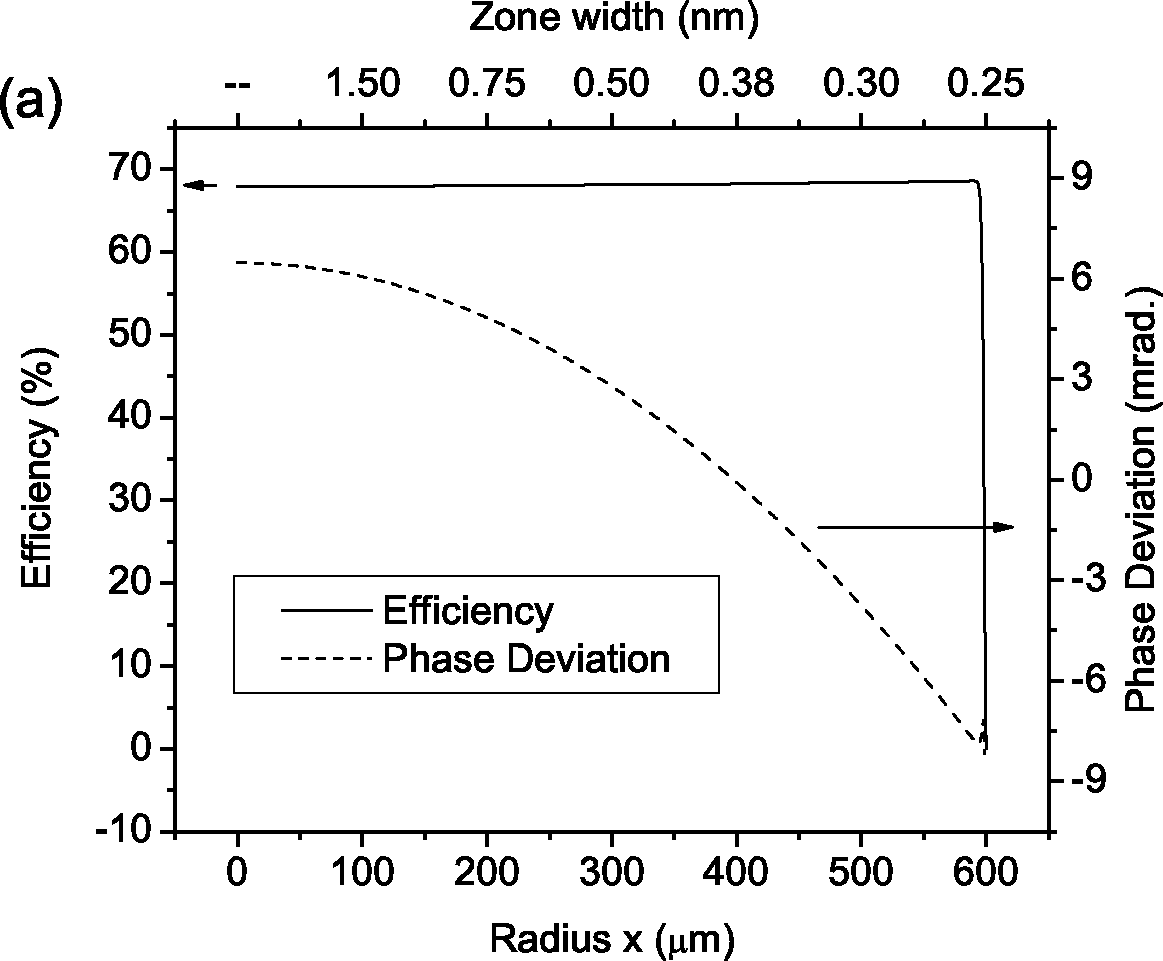}
		}
			\subfigure
		{
		\label{fig:FIG10:b}
		\includegraphics[width=3.2in]{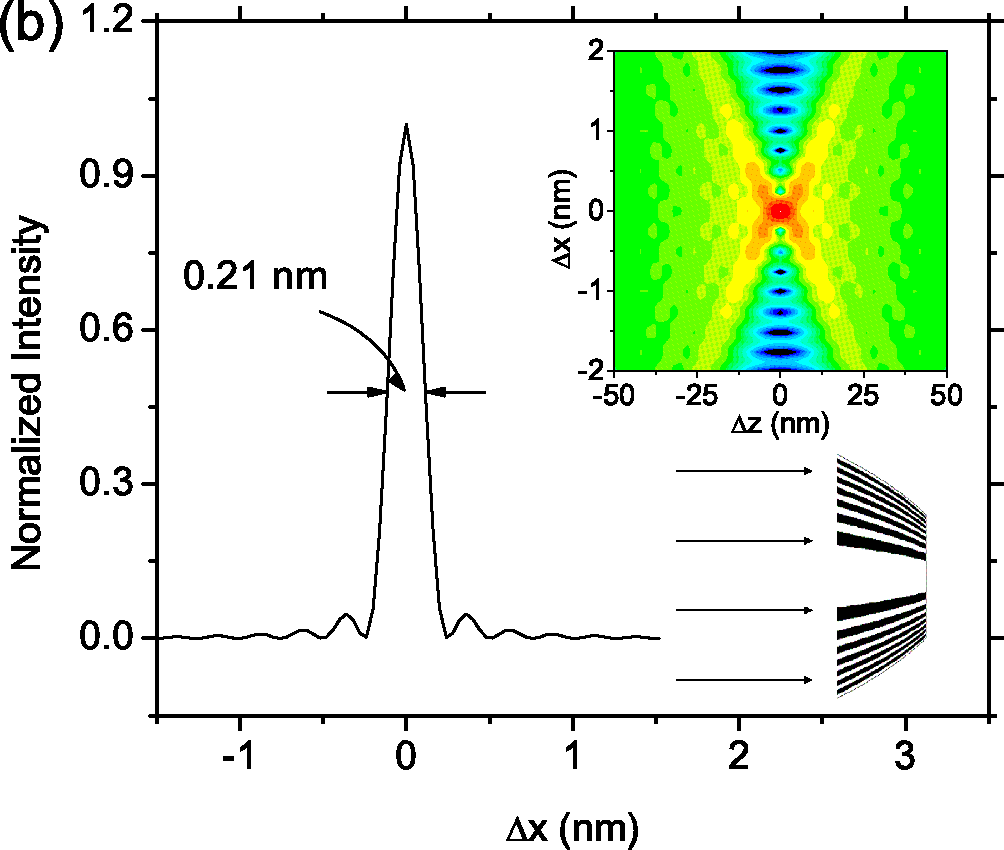}
		}
	\caption{The radial intensity distribution of the $-1^{st}$ order diffraction on the exit surface of an MLL with ideal parabolic zones and outmost zone width 0.25 nm, based on the 2-beam simulation. The dashed curve represents the phase deviation from a perfect converging spherical wave front. b) The intensity profile on the focal plane, showing a peak with FWHM=0.21 nm. The inset on top is the isophote pattern near focus. Intensity is on a logarithm scale, and the one on bottom is a sketch of the ideal structure}
	\label{fig:FIG10}
\end{figure}

We have shown above that a wedged MLL cannot achieve a focus close to the wavelength. Ideal structures, which not only satisfy Bragg condition everywhere so that the diffraction efficiency is high, but also have a right phase locally so that all diffracted waves interfere constructively at the focus, are needed. The first condition requires the deviation function, $\beta_{h}$, to be zero everywhere. We may consider this condition in reciprocal space with the aid of Ewald sphere. In Fig. \ref{fig:FIG9:a} we show a case in which a plane wave is incident normally on a zone plate with local reciprocal lattice vector of $\vec{\rho}_{h}$. For a Bragg condition to be satisfied at a point ($x,z$) within the MLL, the end point of the vector $\vec{k}_{h}=\vec{k}_{0}+\vec{\rho}_{h}$ should lie on the Ewald sphere, which has a radius $k$ and is centered at ($x,z$), so that $\beta_{h}=0$. In addition, for the diffracted wave to converge to the focus \textbf{F}, all $\vec{k}_{h}$ should point to it. In order to satisfy both conditions, $\vec{k}_{h}$ has to take the form given by, 
\[
\vec{k}_{h}=\frac{-kx}{\sqrt{x^{2}+(f-z)^{2}}}\vec{e}_{x}+\frac{k(f-z)}{\sqrt{x^{2}+(f-z)^{2}}}\vec{e}_{x}.
\]
Because $\vec{k}_{0}=k\vec{e}_{z}$, $\vec{\rho}_{h}$ can be determined unambiguously from the relationship $\vec{k}_{h}=\vec{k}_{0}+\vec{\rho}_{h}$. Then, according to the relationship, $\vec{\rho}_{h}=\nabla\phi_{h}$ , we can obtain $\phi_{h}$ by integrating $\vec{\rho}_{h}$. We then obtain the following equation, 
\[
\phi_{h}=-k[\sqrt{x^{2}+(f-z)^{2}}+z+\text{const}].
\]
The integration constant in $\phi_{h}$ can be arbitrary, because it does not affect the Bragg condition. This constant is determined by the second condition: all diffracted waves should interfere constructively at the focus \textbf{F}. By calculating the optical path, we obtain,
\begin{equation}
	\phi_{h}=-k[\sqrt{x^{2}+(f-z)^{2}}-(f-z)].
	\label{eq25}
\end{equation}
We can see that only the $-1^{st}$ order diffraction ($h=-1$) can satisfy both conditions. From Eq. (\ref{eq25}) the ideal zone plate law for an incident plane wave can be deduced,
\begin{equation}
	x^{2}_{n}=n\lambda (f-z)+n^{2}\lambda^{2}/4,
	\label{eq26}
\end{equation}
which is a set of confocal paraboloids [Fig. \ref{fig:FIG9:b}]. Substituting Eq. (\ref{eq26}) into Eq. (\ref{eq12}), with the two-beam approximation we can simplify the system to two partial differential equations,
\begin{equation}
	\left\{\begin{array}{ll}\frac{\partial}{\partial z}E_{0}=i\frac{\pi}{\lambda}(\chi_{0}E_{0}+\chi_{1}E_{-1})\\ \frac{1}{\sqrt{x^{2}+(f-z)^{2}}}[-x\frac{\partial}{\partial x}+(f-z)\frac{\partial}{\partial z}]E_{-1}=i\frac{\pi}{\lambda}(\chi_{-1}E_{0}+\chi_{0}E_{-1})\end{array}\right..
	\label{eq27}
\end{equation}
Solved numerically, the efficiency and the phase deviation of the $-1^{st}$ order diffraction that we obtain are shown in Fig. \ref{fig:FIG10:a}. High diffraction efficiency, 67\%, is observed over all radial positions, and the phase deviation is rather small, indicating that a converging spherical wave front with very small distortion forms on the exit surface of such an MLL. The abrupt drop of the diffraction intensity close to the outmost boundary is a result of incomplete structure beyond it. In Fig. \ref{fig:FIG10:b} we show the intensity profile on the best focal plane with a FWHM of 0.21 nm\footnote{This number is slightly smaller than the conventional diffraction limit, 0.5$\lambda$/NA=0.25 nm from Rayleigh criterion (1-D), because in Rayleigh criterion the resolution is defined as the distance from the peak position to its first minimum. Here we use FWHM, which will be slightly smaller.}, and the inset on top shows the isophotes near the focus. Comparing to the peak half width of 0.34 nm obtained from a wedged MLL with a same outmost zone width of 0.25 nm, we conclude that curved zone profiles are needed to achieve a resolution approaching the wavelength.

\begin{figure}[tbp]
	\centering
		\subfigure
		{
		\label{fig:FIG11:a}
		\includegraphics[width=3.2in]{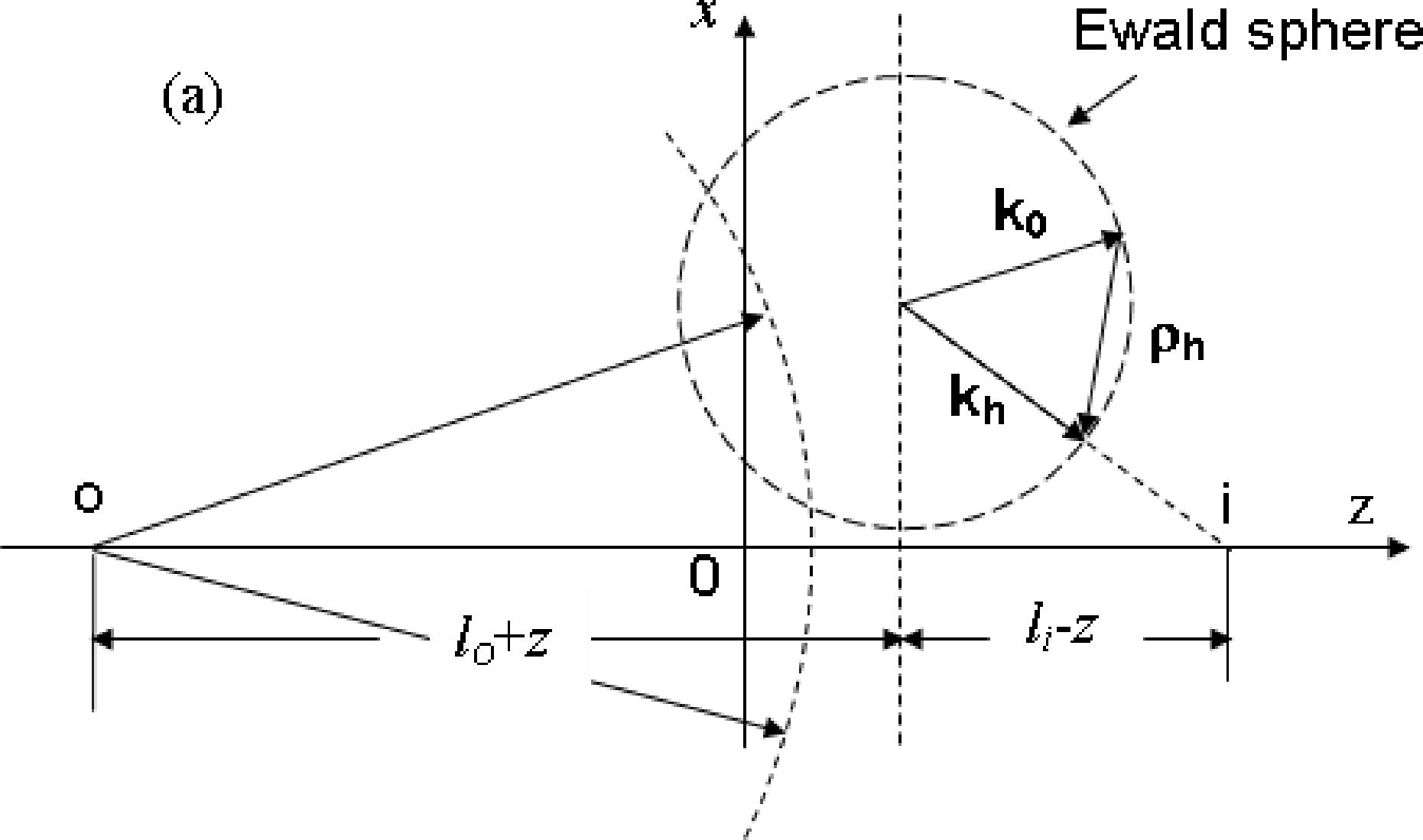}
		}
			\subfigure
		{
		\label{fig:FIG11:b}
		\includegraphics[width=3.2in]{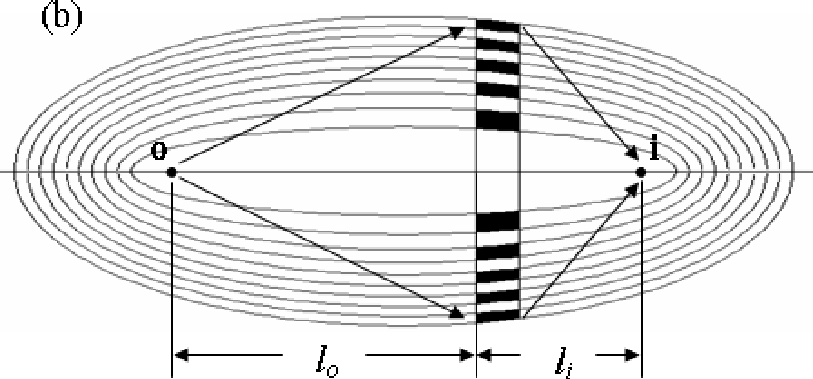}
		}
	\caption{a) A schematic of satisfying Bragg's law everywhere for an incident spherical wave. All diffracted wave vectors point to the focal point \textbf{F}. b) A structure consisting of a set of confocal ellipsoids, serving as an ideal structure to focus a point in object space into a point in image space.}
	\label{fig:FIG11}
\end{figure}

In a similar way, the ideal structure to focus a spherical wave (from a point source) can be obtained as well [see Fig. \ref{fig:FIG11:a}]. The difference is that in this case the incident wave vector is a function of position too,
\[
\vec{k}_{0}=\frac{kx}{\sqrt{x^{2}+(l_{o}+z)^{2}}}\vec{e}_{x}+\frac{k(l_{o}+z)}{\sqrt{x^{2}+(l_{o}+z)^{2}}}\vec{e}_{z},
\]
where $l_{o}$ is the distance from the point source to the entrance surface of the zone plate. To focus it to a point located at a distance $l_{i}$ away from the entrance surface on the downstream side, we find,
\begin{equation}
	\phi_{h}=-k[\sqrt{x^{2}+(l_{o}+z)^{2}}+\sqrt{x^{2}+(l_{i}-z)^{2}}-(l_{o}+l_{i})],
	\label{eq28}
\end{equation}
then we obtain the zone plate law for focusing a spherical wave,
\begin{equation}
	\frac{4x^{2}_{n}}{n^{2}\lambda^{2}/4+n\lambda (l_{o}+l_{i})}+\frac{4[z+(l_{o}-l_{i})/2]^{2}}{(n\lambda/2+l_{o}+l_{i})^2}=1.
	\label{eq29}
\end{equation}
Again, only for $h=-1$ can all requirements be satisfied. Eq. (\ref{eq29}) describes a set of confocal ellipsoids, as shown in Fig. \ref{fig:FIG11:b}. 

\begin{figure}[tbp]
	\centering	
	\includegraphics[width=3.2in]{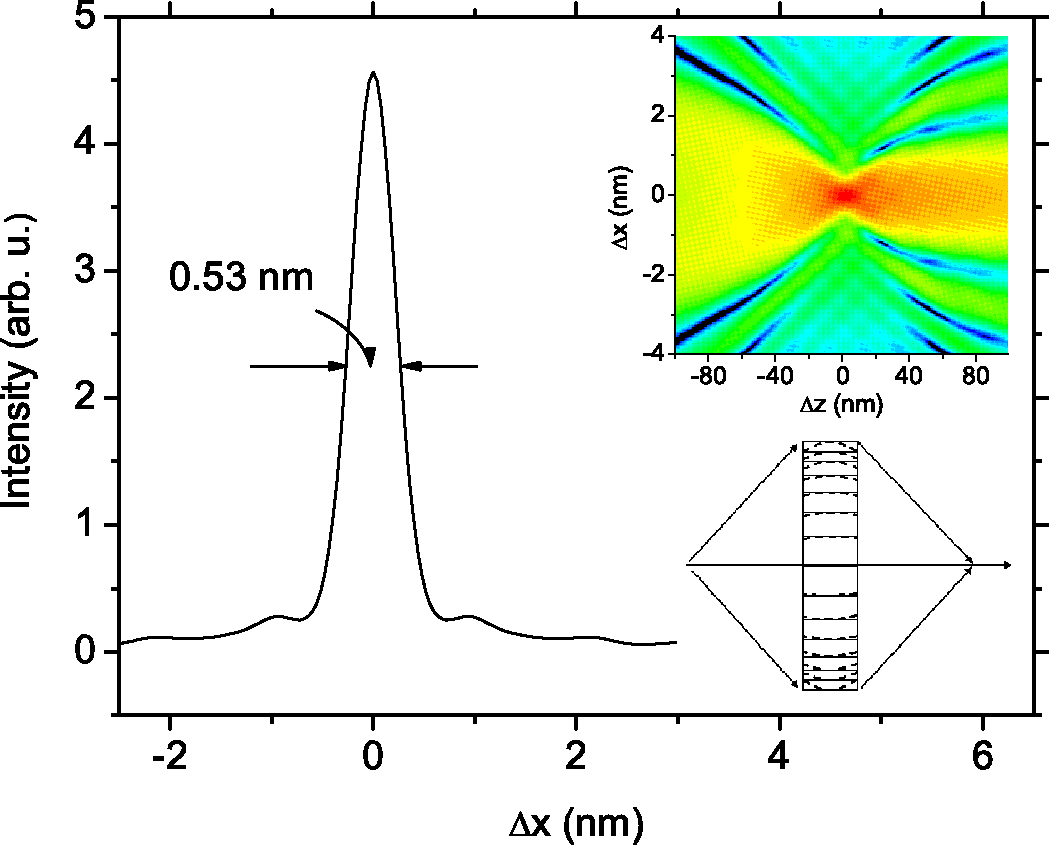}	
	\caption{The focused beam profile on the plane of best focus and the isophotes near the image point (the insert on top) for 1:1 imaging. The insert on bottom shows the difference of an MLL with flat zones (solid line) and an ideal MLL (dashed line).}
	\label{fig:FIG12}
\end{figure}

It is interesting to consider 1:1 imaging in this case. If $l_{o}+l_{i}=4f$ and $l_{o}-l_{i}=-w$, Eq. (\ref{eq29}) then becomes,
\begin{equation}
	x^{2}_{n}=n\lambda f+\frac{n^{2}\lambda^{2}}{16}-\frac{n^{2}\lambda^{2}/4+4 n \lambda f}{(n\lambda/2+4 f)^{2}}(z-w/2)^{2},\qquad 0\leq z\leq w.
	\label{eq30}
\end{equation}
If the thickness $w$ is small enough so that the last term can be neglected, we obtain,
\begin{equation}
	x^{2}_{n}=n\lambda f+\frac{n^{2}\lambda^{2}}{16},
	\label{eq31}
\end{equation}
which is the conventional zone plate law for 1:1 imaging\cite{Kirtz1974}. The inset on bottom of Fig. \ref{fig:FIG12} shows the zone plate structure that satisfies Eq. (\ref{eq31}) (solid flat lines), and the ideal one according to Eq. (\ref{eq30}) (dashed curved lines) for 1:1 imaging. In this case, the flat zones show an increased deviation from the ideal curved ones with increasing radius, indicating there may exist a theoretical limit preventing the NA from approaching unity even for 1:1 imaging. To investigate this limit, we studied the imaging property of a flat MLL when illuminated by a spherical wave emitted from a point source $2f$ away from its center (Fig. \ref{fig:FIG12}). The parameters for the MLL are: $w=13.5\ \mu$m, $f=0.472$ mm, $x_{max}=75\ \mu$m and the outmost zone width of 0.2 nm. It is easy to derive the phase function for this MLL structure from Eq. (\ref{eq31}),
\[
\phi_{h}=4hk(\sqrt{x^{2}/4+f^{2}}-f).
\]
By solving Eq. (\ref{eq12}), we are able to simulate the isophotes near the image point (Fig. \ref{fig:FIG12}). We found a peak FWHM of 0.53 nm, 33\% larger than the diffraction limit from Rayleigh criterion, 0.4 nm. The larger focused beam size, compared to the diffraction limit, is a result of deviating from the ideal structure. From the drawing of MLL structures in Fig. \ref{fig:FIG12}, as the radius increases the flat zone profile deviates further from the ideal elliptical shape. Therefore, as opposed to the conclusion made by Pfeiffer et al.\cite{Pfeiffer2006}, we argue that according to our calculation there is a resolution limit even for 1:1 imaging when an MLL with flat zone profiles is used.

\section{SUMMARIES AND CONCLUSIONS}
In summary, we present a formalism of dynamical diffraction from Fresnel zone plates analogous to the Takagi-Taupin equations for a strained single crystal. The basic equations for dynamical diffraction are derived and employed to study diffraction properties of Multilayer-Laue-Lense with various types of zone profiles, including flat, tilted, wedged and ideally curved. Our study shows that for a thick MLL made of flat zones, in case of plane wave illumination with normal inclination angle, the dynamical diffraction effect of the zone plate would prevent it from focusing the beam size down to a few nanometers. A rough estimation in this case shows that when the thickness is optimized for dynamical diffraction, the achievable NA is about $\sqrt{2\Delta n}$, a result similar to that for waveguides, and in an agreement with the conclusion obtained recently by Pfeiffer et al\cite{Pfeiffer2006}. We also demonstrate that the achievable NA of an MLL with flat zones can be increased by a factor of roughly 2 by tilting zones. The MLL with wedged zones, which satisfy locally the Bragg condition, is studied as well. It is observed that wedged MLL's can reach a NA of $\sim$0.1 when high diffraction efficiency is achieved, i.e., effectively focusing hard x-rays down to several angstroms. Ideal structures which satisfy both the Bragg condition and the phase requirement are obtained with the aid of an Ewald sphere construction. Not unexpectedly, they turn out to be confocal paraboloids for an incident plane wave and confocal ellipsoids for an incident spherical wave.

\section{ACKNOWLEDGEMENTS}
This work supported under contract number DE-AC-02-06CH11357 between UChicago Argonne, LLC and the Department of Energy.

\bibliographystyle{apsrev}

\newpage



\clearpage
\end{document}